\documentstyle[12pt,aaspp4,flushrt]{article}
 
\begin{document}

\baselineskip 14pt
\parskip 4pt
\def\capt{\small \baselineskip 12pt }


\def\subsubsection#1{\smallskip\noindent{\bf #1}\ }

\def\be{\begin{equation}} \def\ee{\end{equation}}

\def\se#1{\S~\ref{sec:#1}}
\def\ses#1{\S\S~\ref{sec:#1}}  \def\sess#1{\ref{sec:#1}}

\def\page{\vfill\eject}  \def\endpage{\vfill\eject}
\def\cl{\centerline} \def\bk{\hfill\break} \def\no{\noindent}

\def\etal{{\it et al.\ }} \def\rms{{\it rms\ }}
\def\cf{{\it cf.\ }} \def\eg{{\it e.g.}} \def\ie{{\it i.e.}}

\def\ltsima{$\; \buildrel < \over \sim \;$}
\def\lsim{\lower.5ex\hbox{\ltsima}}
\def\gtsima{$\; \buildrel > \over \sim \;$}
\def\gsim{\lower.5ex\hbox{\gtsima}}

\def\ifm#1{\relax\ifmmode#1\else$\mathsurround=0pt #1$\fi}
\def\kms{\,{\rm km\,s\ifm{^{-1}}}}
\def\hmpc{\,h\ifm{^{-1}}{\rm Mpc}}
\def\kmsmpc{\,{\rm km\,s\ifm{^{-1}}Mpc^\ifm{{-1}}}}
\def\dd{{d}}

\def\pa{\partial}
\def\phig{\Phi_g} \def\phiv{\Phi_v}

\def\pdf{PDF} \def\ipdf{IPDF}
\def\ps{PS}
\def\pk{$P(k)$}
\def\dm{DM}
\def\cdm{CDM} \def\CDM{CDM} \def\hdm{HDM} \def\HDM{HDM}
\def\mdm{MDM} \def\chdm{CHDM} \def\CHDM{CHDM}
\def\tcdm{TCDM} \def\TCDM{TCDM} \def\pib{PIB}
\def\lcdm{$\Lambda$CDM} \def\scdm{sCDM}
\def\COBE{COBE} \def\cobe{COBE} \def\cmb{CMB}
\def\pot{POTENT} \def\mfpot{MFPOT}
\def\iras{IRAS}
\def\zoa{ZOA} \def\los{LOS}
\def\gi{GI}
\def\lss{LSS}

\def\vp{v_{\scriptscriptstyle P}}
\def\vi{v_{\scriptscriptstyle I}}
\def\vvp{\pmb{$\vp$}}
\def\vvi{\pmb{$\vi$}}
\def\delp{\delta_{\scriptscriptstyle P}}
\def\delg{\delta_{\scriptscriptstyle G}}
\def\deli{\delta_{\scriptscriptstyle I}}
\def\delpi{\delta_{\scriptscriptstyle P(I)}}
\def\bi{b_{\scriptscriptstyle I}}
\def\bo{b_{\scriptscriptstyle O}}
\def\bc{b_{\scriptscriptstyle C}}
\def\betai{\beta_{\scriptscriptstyle I}}
\def\betao{\beta_{\scriptscriptstyle O}}
\def\betac{\beta_{\scriptscriptstyle C}}
\def\mc{MC}
\def\dns{$D_n\!\!-\!\!\sigma$}
\def\sun{\odot}
\def\msun{{\rm M}_{\odot}}
\def\lsun{{\rm L}_{\odot}}
\def\omt{\Omega_{tot}}
\def\omo{\Omega_o}
\def\omm{\Omega}
\def\oml{\Omega_\Lambda}
\def\omb{\Omega_b}
\def\omk{\Omega_k}
\def\oma{\Omega_a}
%
\def\pmb#1{\setbox0=\hbox{#1}%
 \kern-.025em\copy0\kern-\wd0
 \kern.05em\copy0\kern-\wd0
 \kern-.025em\raise.0433em\box0}
\def\vv{\pmb{$v$}}
\def\vq{\pmb{$q$}}
\def\vx{\pmb{$x$}}
\def\vd{\pmb{$d$}}
\def\vy{\pmb{$y$}}
\def\vr{\pmb{$r$}}
\def\vg{\pmb{$g$}}
\def\vk{\pmb{$k$}}
\def\vz{\pmb{$z$}}
\def\vs{\pmb{$s$}}
\def\vV{\pmb{$V$}}
\def\vB{\pmb{$B$}}
\def\vF{\pmb{$F$}}
\def\vR{\pmb{$R$}}
\def\vS{\pmb{$S$}}
\def\vL{\pmb{$L$}}
\def\vC{\pmb{$C$}}
\def\vpsi{\pmb{$\psi$}}
\def\veps{\pmb{$\epsilon$}}
\def\vth{\pmb{$\th$}}
\def\vnabla{\pmb{$\nabla$}}
\def\div{\vnabla\!\cdot\!}
\def\rot{\vnabla\!\times\!}
\def\rhat{\hat r}
\def\divv{\div\vv}
\def\rotv{\rot\vv}
\def\delc{\delta_c}
\def\deld{\delta_d}
\def\del0{\delta_0}
\def\del1{\delta_1}
\def\sigm{\sigma_m}

\title{COSMOLOGICAL IMPLICATIONS OF LARGE-SCALE FLOWS\footnote{in 
``Galaxy Scaling Relations: Origins, Evolution and Applications", 
ed. L. da Costa (Springer) in press (1997). A more complete
discussion is in ``Formation of Structure in the Universe", eds. A. Dekel
\& J.P. Ostriker (Cambridge University Press) in press (1997).} }

\author{\bf Avishai Dekel}
\bigskip

\affil{Racah Institute of Physics, The Hebrew University, Jerusalem 91904, 
Israel \\ and University of California, Berkeley \& Santa Cruz}


\begin{abstract}

Cosmological implications of the observed large-scale peculiar velocities
are reviewed, alone or combined with redshift surveys and CMB data. 
The latest version of the POTENT method for reconstructing 
the underlying three-dimensional velocity and mass-density fields is described. 
The initial fluctuations and the nature of the dark matter are addressed via
statistics such as bulk flow and mass power spectrum.
The focus is on constraining the mass density parameter $\Omega$,
directly or via the parameter $\beta$ which involves the unknown relation
between galaxies and mass.
The acceptable range for $\Omega$ is found to be $0.4-1.0$.
The range of $\beta$ estimates is likely to reflect non-trivial
features in the galaxy biasing scheme, such as scale dependence.
Similar constraints on $\Omega$ and $\Lambda$
from global measures are summarized.

\end{abstract}

\section{INTRODUCTION}
\label{sec:de_intro}

A major goal of the analysis of cosmic flows is measuring the cosmological
parameters, in particular the mass density parameter, $\Omega$ (or $\omm$).
As illustrated in Figure~\ref{fig:de_omega},
the data can be used to constrain $\Omega$
in several different ways. 
\begin{figure}[th]
\vspace{5.5truecm}
\includegraphics{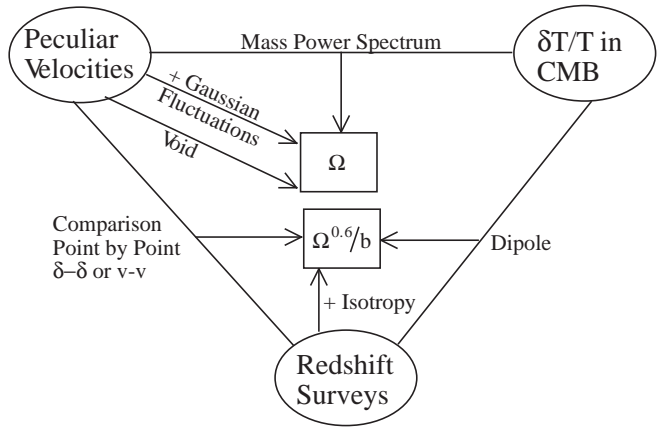}
\caption{\capt  
Methods for measuring $\Omega$ and $\beta$ from large-scale flows.
}
\label{fig:de_omega}
\end{figure}
Methods that are based on the peculiar velocity data alone
(\se{de_omega})
are independent of the ``biasing" relation between galaxy and mass density.
They refer to the present-day large-scale structure
(\lss) which is insensitive to the cosmological constant $\Lambda$. 
They can thus serve to measure 
$\Omega$ directly, but they involve relatively large errors.
These methods largely rely on the assumption (supported by observations,
\eg, Nusser, Dekel \& Yahil 1995)
that the initial fluctuations were drawn from a {\it Gaussian} random field.

All the methods that use the spatial
distribution of galaxies must depend on the biasing relation between
the densities of galaxies and mass.
In the linear approximation to gravitational instability theory (\gi), 
which is roughly valid for the fields when they are smoothed on very large
scales, what is actually being measured by these methods is the degenerate
parameter $\beta\equiv \Omega^{0.6}/b$ rather than $\Omega$ itself,
where the density fluctuations of galaxies and mass are assumed to be related
via a linear biasing relation, $\delg =b \delta$.
These methods include, for example,
measurements of redshift-space distortions from redshift surveys
under the assumption of global isotropy,  
comparisons of the galaxy distribution with the CMB dipole,
and comparisons of the observed peculiar velocities with the
galaxy distribution (or the predicted velocities) deduced from 
redshift surveys (\se{de_beta}).
The most recent best estimates of $\beta$ for IRAS galaxies
lie in the range $0.5 \leq \betai \leq 0.9$.
As argued below, the contamination by non-trivial biasing
introduces a significant uncertainty in the translation of the various $\beta$
estimates to a reliable measurement of $\Omega$.

Another way of determining $\Omega$ is by comparing the mass power
spectrum from observed peculiar velocities with the spectrum of
fluctuations in the \cmb\ (\se{de_fluct_cobe}, \se{de_fluct_cmb}).
This comparison involves more detailed modeling of the formation of structure,
and the current studies are limited to the \cdm\ family of models, 
allowing for a non-zero cosmological constant, a
possible tilt in the spectrum with or without tensor fluctuations,
and a mixture of dark-matter species.

The outline is as follows:
\se{de_rec} 
briefly describes reconstruction methods from peculiar velocities.
\se{de_fluct}
discusses the statistics of mass-density fluctuations, including 
bulk flow and power spectrum,
as determined by the velocity data alone or combined with CMB data.  
\se{de_omega}
focuses on direct estimates of $\Omega$ from peculiar velocities alone.
\se{de_beta}
presents estimates of $\beta$ by comparing velocity
   and galaxy density data, and addresses the issue of biasing.
\se{de_param} 
puts the results in perspective by summarizing
other measures of the cosmological parameters.
\se{de_conclusion}
provides a summary of the results.

\section {POTENT RECONSTRUCTION FROM PECULIAR VELOCITIES}
\label{sec:de_rec}

\subsection {Data for Velocity Analysis}
\label{sec:de_rec_data}
 
The key for measuring peculiar velocities are the distance indicators
(for review: Willick 1997).
The current typical intrinsic scattering the Tully-Fisher (TF) distance
indicator is at best $\sigm\!\sim\!0.33$ mag, corresponding to a relative
distance error of $\Delta\!=(\ln10/5)\sigm\!\approx\!0.15$.
 
The most comprehensive catalog of peculiar velocity data available today
is the Mark III catalog
(Willick \etal 1995; 1996; 1997a), which is a careful
compilation of several data sets under the assumption that all galaxies
trace the same underlying velocity field.
The merger was non-trivial because
the observers differ in their selection
procedure, the quantities they measure, the method of measurement and
the TF calibration techniques.
 
The original Mark II catalog, which was used
in the first application of POTENT (Dekel \etal 1990;
Bertschinger \etal 1990),
consisted of about 1000 galaxies
(mostly Lynden-Bell \etal 1988 and Aaronson \etal 1982).
The extended Mark III catalog consists of $\sim 3400$ galaxies
(dominated by Mathewson \etal 1992).
This sample enables a reasonable recovery of the dynamical fields with
$12\hmpc$ smoothing in a sphere of radius $\sim\!60\hmpc$ about
the Local Group (LG), 
extending to $\sim\!80\hmpc$ in certain regions (\se{de_rec_pot_maps}).
More uniformly sampled data of more than 1000 spiral galaxies
in the north is in preparation, and was already subject to
preliminary analysis (SFI, Giovanelli \etal 1997).

\subsection {Methods of Velocity Analysis}
\label{sec:de_rec_methods}

One way of classifying the methods of velocity analysis is as follows:

\medskip

\moveright 1.6cm
\vbox{\offinterlineskip
\hrule
\halign{&\vrule#&
        \strut\quad\hfil#\quad\cr
height2pt&\omit&&\omit&&\omit&\cr
& &&Inferred Distance Space&&Redshift-Space + V Model&\cr
\noalign{\hrule}
&Forward\hfil&&POTENT\hfil&&VELMOD\hfil&\cr
&TF\hfil     &&{\it selection + distance bias}\hfil&&{\it selection bias}
  \hfil&\cr
\noalign{\hrule}
&Inverse\hfil&&POTINV\hfil&&MFPOT\hfil&\cr
&TF\hfil     &&{\it distance bias}\hfil&&{\it z-space smoothing}
\hfil&\cr
}
\hrule}

\medskip


\no
The ``forward" and ``inverse" methods refer to whether the TF relation
is interpreted as $M(\eta)$ or $\eta(M)$ ($M$ being the absolute magnitude
and $\eta$ the rotation velocity). 
The difference is crucial because
the apparent magnitude depends on distance 
while $\eta$ is not, and because the selection depends on magnitude and is 
independent of $\eta$.  On the other hand,
the velocity field can be computed either from peculiar velocities that
were evaluated at each galaxy's TF-inferred position, $\vd$,
or by fitting a parametric model for the potential
(and thus the velocity) field in redshift space, $\vz$.
Each method is affected by different systematic errors,
and techniques have been developed for statistically correcting them.
The success of each technique has been tested using mock catalogs,
and the goal is to have the different methods recover consistent
results.

The original POTENT described below
is a {\it forward} TF method in d-space.
It has to deal with {\it selection} bias of the TF parameters because
of the magnitude limit, and with {\it Malmquist} bias in the inferred
distances and velocities arising from the random distance errors 
convolved with the geometry of space and the clumpiness of the galaxy
distribution.  Methods have been developed for statistically
correcting these biases. In particular, the correction of the Malmquist 
bias requires external information about the underlying number density 
of galaxies in the samples from which galaxies were selected for the 
peculiar velocity catalogs (see Willick 1997).
 
One can alternatively infer distances using the {\it inverse} TF
relation. This eliminates the
selection bias, but there is still an inferred-distance Malmquist bias.
In the inverse-TF case, the distance bias can in principle be corrected using 
information that is fully contained in the catalog itself 
(Landy \& Szalay 1992).
In practice, the quality of the correction is limited by the
sparseness of the sampling.
The POTENT analysis of the inverse data corrected this way
is termed POTINV 
(Eldar, Dekel \& Willick 1997).

If the selection does not explicitly depend on $\eta$, Malmquist bias
can be eliminated by minimizing $\eta$ residuals in redshift space
without ever inferring actual distances to individual galaxies.
The distance is replaced by $r=z-u_\alpha(z)$, where $u_\alpha$ 
is a parametric model for the radial peculiar velocity field.
If the forward TF relation is used, as in VELMOD (Willick \etal 1997b;
\se{de_beta_velmod}),
the method still has to correct for selection bias.
The use of the {\it inverse} TF relation (Schechter 1980)
guarantees in this case that
both the selection bias and the distance bias are eliminated, at the
expense of over-smoothing due to the representation of the 
fields in redshift space.
Recent implementations of such methods were termed MFPOT 
(Blumenthal, Dekel \& Yahil 1997)
and ITF (Davis, Nusser, \& Willick 1996).

Another way of distinguishing between the methods is by their goals.
The methods working in d-space can serve for reconstruction of 3D
maps of the velocity and mass-density fields, unbiased and
uniformly smoothed with equal-volume weighting throughout the volume. 
These fields can then be straightforwardly compared to other data and to theory
in order to obtain cosmological implications (\eg, \se{de_beta_dd}).
Alternatively, one may direct the method
to estimating certain parameters of the model (\eg, $\beta$)
without ever reconstructing uniform maps. The
redshift-space methods serve this purpose well (\eg, \se{de_beta_velmod}). 

Yet another characteristic of some of the methods is the usage of a whole-sky 
redshift survey (such as IRAS 1.2 Jy) as an intrinsic part of the 
reconstruction from peculiar velocities. This is the case in the 
SIMPOT, VELMOD and ITF methods (\se{de_beta}). These methods are geared
towards determining $\beta$, with SIMPOT also providing uniform 
reconstruction maps.

Finally, one can focus on optimal formal treatment of the random errors,
which are in fact the main obstacle. A method based on Wiener Filtering has
been developed for recovering the most probable mean field
from the noisy peculiar-velocity data in d-space
(Zaroubi, Dekel \& Hoffman, in preparation). 
This can serve as a basis for constrained 
realizations of uniform smoothing, each of which being an equally good 
guess for the structure in our real cosmological neighborhood.


\subsection {Correcting Malmquist Bias}
\label{sec:de_rec_pot_im}

The selection bias in the calibration of the forward TF relation 
can be corrected once the selection function is known
(see Willick 1994).
But then,
the TF inferred distance, $d$, and the mean peculiar velocity at a given
$d$, suffer from a {\it Malmquist} or {\it inferred-distance} bias. 
The distances, either forward or inverse,
are corrected for Malmquist bias in a statistical way before being fed as
input to POTENT-like procedures.

If $M$ is distributed normally for a given $\eta$, with standard deviation 
$\sigm$, then the forward inferred distance $d$ 
of a galaxy at a true distance $r$
is distributed log-normally about $r$, with relative error
$\Delta\!\approx\! 0.46\sigm$.
Given $d$, the expectation value of $r$ is
\begin{equation}
E(r \vert d)=
{ \int_0^\infty r P(r\vert d) {\rm d}r
\over
\int_0^\infty P(r\vert d) {\rm d}r } =
{ \int_0^\infty r^3 n(r)\ {\rm exp}
\left( -{[{\rm ln}(r/d)]^2 \over 2\Delta^2} \right) {\rm d}r
\over
\int_0^\infty r^2 n(r)\ {\rm exp}
\left( -{[{\rm ln}(r/d)]^2 \over 2\Delta^2} \right) {\rm d}r } \ ,
\label{eq:de_malmquist}
\end{equation}
where $n(r)$ is the number density in the underlying distribution from
which galaxies were selected.
The deviation of $E(r\vert d)$ from $d$ reflects the bias.
The homogeneous part arises from the geometry of space ---
the inferred distance $d$ underestimates $r$
because it is more likely to have been scattered by errors from $r>d$
than from $r<d$, the volume being $\propto r^2$.
If $n\!=\!const$, equation \ref{eq:de_malmquist} reduces to
$E(r\vert d) = d e^{3.5 \Delta^2}$,
in which the distances should simply be multiplied by a
factor, 8\% for $\Delta=0.15$, equivalent to changing the
zero-point of the TF relation.
Fluctuations in $n(r)$ are responsible for the inhomogeneous bias
(IM), which  systematically enhances the inferred
density perturbations and thus
the value of $\Omega$ inferred from them.  

 In one version of the Mark III data for POTENT analysis, 
the forward IM bias is corrected in two steps.
First, the galaxies are grouped in $z$-space
(Willick \etal 1995), reducing the distance error of each group of $N$
members to $\Delta/\sqrt N$ and thus significantly weakening the bias.
With or without grouping, the noisy inferred distance of each object,
$d$, is replaced by $E(r\vert d)$, with an assumed 
$n(r)$ properly corrected for grouping if necessary.  
This procedure has been tested using realistic mock data from
N-body simulations,
showing that IM bias can be reduced to the level of a few percent.  
The practical uncertainty
is in $n(r)$, which can be approximated for example by the high-resolution
density field of IRAS or optical galaxies, 
or by the recovered
mass-density itself in an iterative procedure under certain assumptions about
how galaxies trace mass.  The resultant correction to the density 
recovered by POTENT is $<\!20\%$ even at the highest peaks.

\def\dinv{d}
\def\Dinv{\Delta}

Distances are alternatively inferred via the {\it inverse} 
TF relation between internal velocity parameter 
$\eta$ and magnitude $m$, $\eta = \eta^0 (m-5 \log \dinv)$.
Under the assumption that the selection was independent of $\eta$ and was
not an explicit function of distance,
the expectation value of the true distance $r$ given $\dinv$ is
\begin{equation}
E(r|\dinv ) = \dinv\, e^{3 \Dinv^2 /2}\,
{f(\dinv e^{\Dinv^2}) /f(\dinv ) } \ ,
\label{eq:de_dinv}
\end{equation}
where $\Dinv \equiv (\ln 10 / 5){\sigma_\eta / \eta^0}$.
In this case, the required density function, $f(\dinv)$, is in $\dinv$-space,
and is derivable from the sample itself (Landy \& Szalay 1992).
Eldar, Dekel \& Willick (1997) have applied this correction
to the inverse distances in the Mark III catalog, to serve as input
for a POTENT analysis (POTINV). The agreement between the forward POTENT
and POTINV results are well within the level of the random errors.

\subsection{Smoothing the Radial Velocities}
\label{sec:de_rec_pot_twf}

The goal of the POTENT analysis is to recover from the collection
of Malmquist-corrected, radial peculiar velocities $u_i$ 
at inferred positions $\vd_i$ the underlying 3D velocity field $\vv(\vx)$
and the associated mass-density fluctuation field $\delta(\vx)$, 
smoothed with a Gaussian of radius $R_s$ (we denote hereafter
a 3D Gaussian window of radius $12\hmpc$ by G12, etc.). 
The first, most difficult step is the smoothing, or  
interpolation, into a radial velocity field with minimum bias, $u(\vx)$.
The desire is to reproduce the $u(\vx)$
that would have been obtained had the true $\vv(\vx)$ been sampled
densely and uniformly and smoothed with a spherical
Gaussian window of radius $R_s$.  
With the data as available, $u(\vx_c)$ is taken to be the value at
$\vx\!=\!\vx_c$ of an appropriate {\it local} velocity model
$\vv(\alpha_k,\vx\!-\!\vx_c)$. The model parameters $\alpha_k$ are
obtained by minimizing the weighted sum of residuals,
\begin{equation}
S = \sum_i W_i\,
[u_i-\hat{\vx}_i\cdot\vv(\alpha_k,\vx _i)]^2 \ ,
\label{eq:de_sumv}
\end{equation}
within an appropriate local window $W_i\!=\!W(\vx_i,\vx_c)$.
The window is a Gaussian, modified such that it
minimizes the combined effect of the following three types of errors.

\subsubsection{Tensor window bias.}  
Unless $R_s\ll r$, the $u_i$s cannot be
averaged as scalars because the directions $\hat{\vx}_i$ differ from
$\hat{\vx}_c$, so $u(\vx_c)$ requires a
fit of a local 3D model as in Eq. \ref{eq:de_sumv}.  
The original POTENT used the
simplest local model, $\vv(\vx)\!=\!\vB$ of 3 parameters, for which the
solution can be expressed explicitly in terms of a tensor window
function (Dekel \etal 1990).
However, a bias occurs because
the tensorial correction to the spherical window has
conical symmetry, weighting more heavily objects of large
$\hat{\vx}_i\!\cdot\!\hat{\vx}_c$.  
A way to reduce this bias is by generalizing the zeroth-order $\vB$ 
into a 9-parameter first-order velocity model,
$ \vv(\vx)\!=\!\vB\! +\! \bar{\bar{\vL}}\! \cdot\! (\vx\!-\!\vx_c) $,
with $\bar{\bar{\vL}}$ a symmetric tensor that automatically ensures
local irrotationality.  The linear terms tend to ``absorb" most of
the bias, leaving $\vv(\vx_c)\!=\!\vB$ less biased.
Unfortunately, a high-order model tends to
pick undesired small-scale noise.  The optimal compromise for the 
Mark III data was found to be a 9-parameter model fit out
to $r\!=\!40\hmpc$, smoothly changing to a 3-parameter fit beyond
$60\hmpc$ (Dekel \etal 1997).

\subsubsection{Sampling-gradient bias.} 
If the true velocity field is
varying within the effective window, the non-uniform sampling
introduces a bias because the smoothing is galaxy-weighted whereas
the aim is equal-volume weighting.
The simplest way to correct this bias is by weighting each object 
with the local volume that it ``occupies", or the inverse of the local density.
A crude estimate of this volume is $V_i\!\propto\!R_n^3$,
where $R_n$ is the distance to the
n-th neighboring object (\eg, $n\!=\!4$).  This procedure
is found via simulations to reduce
the sampling-gradient bias in Mark III to negligible levels typically
out to $60\hmpc$ as long as one keeps out of the Galactic zone of avoidance.
The $R_n(\vx)$ field can serve later as a flag for poorly sampled regions,
to be excluded from any quantitative analysis.

\subsubsection{Reducing random errors.}
The ideal weighting for reducing the effect of Gaussian noise
has weights $W_i\!\propto\!\sigma_i^{-2}$, where $\sigma_i$ are the
distance errors. Unfortunately, this weighting spoils
the carefully designed volume weighting, biasing $u$
towards its values at smaller $r_i$ and at nearby clusters where the
errors are small.  A successful compromise is to weight by both, \ie
\begin{equation}
W(\vx_i,\vx_c) \propto
V_i\, \sigma_i^{-2}\, \exp [-(\vx_i-\vx_c)^2/ 2R_s^2] \ .
\label{eq:de_window}
\end{equation}

The resultant errors in the recovered
fields are assessed by Monte-Carlo simulations. 
We generate noisy data via full, realistic 
Monte-Carlo mock catalogs, where the noise is added as scatter in the 
TF quantities (Kolatt \etal 1996). The error in the final $\delta$ 
at a grid point is
estimated by the standard deviation of the recovered $\delta$ over the
Monte-Carlo simulations, $\sigma_\delta$ (and similarly $\sigma_v$).  In
the well-sampled regions, which extend in Mark III out to
$40\!-\!60\hmpc$, the errors are
$\sigma_\delta\!\approx\!0.1\!-\!0.3$, but they may blow up in certain regions
at large distances.  To exclude noisy regions, any quantitative analysis
could be limited to points where $\sigma_v$ and $\sigma_\delta$ are
within certain bounds.

\subsection{From Radial Velocity to Density Fields}
\label{sec:de_rec_pot_veldel}

 If the \lss\ evolved according to \gi, then the large-scale velocity field
is expected to be {\it irrotational}, $\rotv\!=\!0$.  Any vorticity mode
would have decayed during the linear regime as the universe expanded, and,
based on Kelvin's circulation theorem, the flow remains vorticity-free in
the mildly-nonlinear regime as long as it is laminar. Bertschinger 
\& Dekel (1989) have demonstrated that irrotationality is valid to 
a good approximation when a nonlinear 
velocity field is properly smoothed over.  Irrotationality
implies that the velocity field can be derived from a scalar potential,
$\vv(\vx)\!=\!-\vnabla\Phi(\vx)$, so the radial velocity field
$u(\vx)$ should contain in principle
enough information for a full 3D reconstruction.  In
the POTENT procedure, the potential is computed by integration along radial
rays from the observer,
\begin{equation}
\Phi(\vx) = -\int_0^r u (r',\theta,\phi) dr' \ .
\label{eq:de_velpot}
\end{equation}
The two missing transverse velocity components are then
recovered by differentiation.


The final step of the POTENT procedure is the derivation of the
mass-density fluctuation field associated with the peculiar
velocity field. This requires a solution to the equations of GI
in the mildly-nonlinear regime with mixed boundary conditions.

Let $\vx, \vv$ be the position and peculiar velocity
in comoving units
(corresponding to $a\vx$ and $a\vv$ in physical units,
with $a(t)$ the universal expansion factor).
Let $\delta\equiv (\rho-\bar\rho)/\rho$
be the mass-density fluctuation.  The equations governing the
evolution of fluctuations of a
pressureless gravitating fluid in a standard cosmological background
during the matter era are the {\it Continuity} equation,
the {\it Euler} equation of motion, and the {\it Poisson} field equation.

In the {\it linear} approximation,
the growing mode of the solution, $\delta \propto D(t)$,
is irrotational and can be expressed in terms of
$f(\Omega)\!\equiv\!H^{-1}\dot{D}/D \!\approx\! \Omega^{0.6}$.
The corresponding linear relation between density and velocity is
$
\del1 = -f^{-1}\divv
$.
The use of $\del1$ is limited to the small
dynamical range between a few tens of megaparsecs
and the $\sim\!100\hmpc$ extent of the current samples.
However, the sampling of galaxies enables
reliable dynamical analysis with a smoothing radius as small as
$\sim\!10\hmpc$, where $\vert\divv\vert$ obtains values 
larger than unity and therefore nonlinear effects play a role.
Even reconstruction with $\sim 5 \hmpc$ smoothing may be feasible
in well-sampled regions nearby.
Figure \ref{fig:de_delta}
shows that $\del1$ becomes a severe underestimate at large $\vert\delta\vert$.  
Mild nonlinear
effects carry crucial information about the formation of \lss, and should
therefore be treated properly.

\begin{figure}[th]
\vspace{7.9truecm}
{\includegraphics{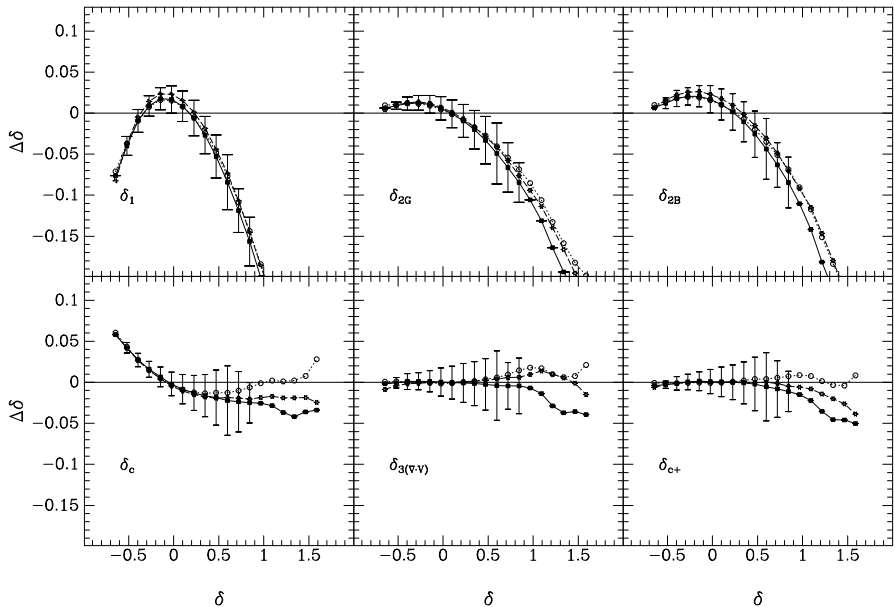}}
\caption{\capt  
Quasi-linear velocity-to-density approximations.
$\Delta\delta\equiv\delta_{approx}(\vv)-\delta_{true}$. The mean and standard
deviation are from large \cdm\ N-body simulations normalized to
$\sigma_8=1$, smoothed G12.
The three curves correspond to different models: standard CDM (squares,
solid), tilted CDM ($n=0.6$, stars, dashed), and open CDM ($\Omega=0.2$,
open circles, dotted) 
(Ganon \etal 1997).
}
\label{fig:de_delta}
\end{figure}

 A basis for useful {\it mildly-nonlinear} relations is provided by the
{\it Zel'dovich} (1970) approximation.  The displacements of particles
from their initial, Lagrangian positions $\vq$ to their Eulerian
positions $\vx$ at time $t$ are assumed to have a universal time
dependence, and thus,
$ \vx(\vq) - \vq = f(\Omega)^{-1} \vv(\vq) $.
For the purpose of approximating
\gi, the Lagrangian Zel'dovich approximation can be interpreted in {\it
Eulerian} space, $\vq(\vx)\!=\!\vx\!-\!f^{-1} \vv(\vx)$, provided that
the flow is {\it laminar} with no orbit mixing, or when multi-streams 
are appropriately smoothed over.  The
solution of the continuity equation then yields (Nusser \etal 1991)
\begin{equation}
\delta_c(\vx) = \Vert I - f^{-1} {\pa \vv / \pa \vx} \Vert -1 \ ,
\label{eq:de_delc}
\end{equation}
where the bars denote the Jacobian determinant and $I$ is the unit
matrix.  The Zel'dovich displacement is first order in $f^{-1}$ and
$\vv$ and therefore the determinant in $\delta_c$ includes
 second- and third-order terms as well,
involving sums of double and triple products of partial derivatives 
which we term $\Delta_2(\vx)$ and $\Delta_3(\vx)$ respectively.

The approximation
$\delta_c$ can be improved by slight adjustments of the coefficients
of the $n$th-order terms, 
\begin{equation}
\delta_{c+} = -(1+\epsilon_1) f^{-1} \divv
             +(1+\epsilon_2) f^{-2} \Delta_2
             +(1+\epsilon_3) f^{-3} \Delta_3 \ .
\label{eq:de_delc+}
\end{equation}
The coefficients were empirically tuned to best fit the CDM simulation
of $12\hmpc$ smoothing over the whole range of $\delta$ values, 
with $\epsilon_1=0.06$, $\epsilon_2=-0.13$ and $\epsilon_3=-0.3$.
This approximation is found to be robust to uncertain features
such as the value of $\Omega$, the shape of the power spectrum, 
and the degree of non-linearity as determined by the fluctuation amplitude 
and the smoothing scale.
Such robustness is crucial when a quasilinear
approximation is used for determining $\Omega$ (\se{de_omega}).
This is the approximation currently used in POTENT.

 Fig. \ref{fig:de_delta} compares the accuracy of the various approximations
using the N-body simulations.
$\delta_c$ is the best among the physically motivated approximations,
which also include two second-order approximations
(Bernardeau 1992; Gramman 1993). The latter do somewhat better at the 
negative tail, but they provide severe underestimates in the positive tail.
$\delta_{c+}$ is an excellent
robust fits over the whole mildly-nonlinear regime. 

We note in passing that the relation \ref{eq:de_delc} is not easily
invertible to solve for $\vv$ when $\delta$ is given, \eg,
from redshift surveys, but
a useful approximation derived from simulations is
$\divv = -f\delta /(1+0.18\delta)$.

\subsection{Testing with Mock Catalogs}
\label{sec:de_rec_pot_mock}

The way to optimize POTENT and other reconstruction methods is
by minimizing the systematic errors when applied to mock catalogs.
It is important that these mock catalogs mimic the real data as closely 
as possible.  Such mock catalogs have been produced, for example, to mimic the 
Mark III and the IRAS 1.2Jy catalogs (Kolatt \etal 1996).
They are publically available and 
serve as standard ``benchmarks" for the competing methods.

The procedure for making these mock catalogs
involves two main steps: a dynamical N-body simulation that mimics our actual
cosmological neighborhood, and the generation of galaxy catalogs from it.

\begin{figure}[th]
\vspace{6.7truecm}
{\includegraphics{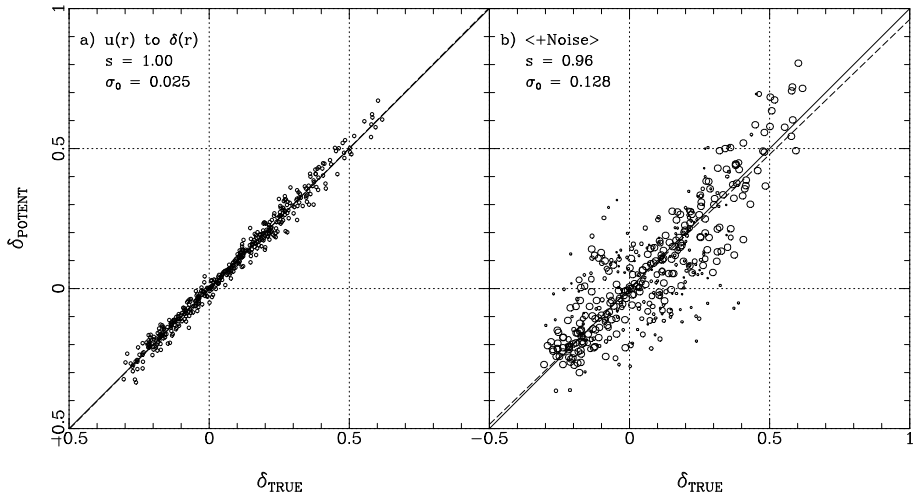}}
\caption{\capt 
Systematic errors in the POTENT analysis.
The density field recovered by POTENT from the mock data
is compared with the ``true" G12 density. 
The comparison is at uniform grid points within a volume of radius $40\hmpc$.
Left: The input to POTENT is the true, G12-smoothed radial velocity. 
The small scatter of 2.5\% reflects small deviations from potential flow, 
the scatter in the non-linear approximation Eq.~\ref{eq:de_delc+},
and numerical errors. 
Right: The input is noisy and sparsely-sampled mock data. Shown is the average
of 10 random realizations. The global bias is only -4\% (Dekel \etal 1997).
}
\label{fig:de_potbias}
\end{figure}

Figure \ref{fig:de_potbias} 
demonstrate the quality of the POTENT reconstruction from the Mark III catalog
by comparing the recovered density field to the true G12-smoothed field.
This comparison is done 
at the points of a uniform grid inside a volume of effective radius $40\hmpc$.
The field shown is
the average of the fields recovered from ten Monte Carlo mock catalogs 
of noisy galaxy velocities sampled sparsely and nonuniformly.
One can see that the remaining systematic errors are small.
The final systematic error is not correlated with the signal 
(slope $\sim$unity in the scatter diagram) and is on the order 
of $\Delta\delta\sim 0.13$.
The random errors are not a major obstacle in certain well-sampled
regions (such as the Great Attractor), but they
become severe in poorly-sampled regions 
(such as parts of the Perseus-Pisces region near the Galactic plane).  
The errors derived from the noisy mock catalogs
are used to eliminate poorly-recovered regions from quantitative
analyses.

\subsection{Maps of Velocity and Density Fields}
\label{sec:de_rec_pot_maps}

Figure \ref{fig:de_pot_map} shows Supergalactic-plane maps of the 
velocity field in the
\cmb\ frame and the associated $\delta_{c+}$ field (for $\Omega\!=\!1$) as
recovered by POTENT from the Mark III catalog.  The recovery is
reliable out to $\sim\!60\hmpc$ in most directions
outside the Galactic plane ($Y\!=\!0$).
Both large-scale ($\sim\!100\hmpc$) and small-scale ($\sim\!10\hmpc$)
features are important; \eg, the bulk velocity reflects
properties of the initial fluctuation power spectrum (\se{de_fluct}),
while the small-scale variations indicate the value of $\Omega$ 
(\se{de_omega}).

 The velocity map shows a clear tendency for motion
from right to left, in the general direction of the LG motion
in the CMB frame ($L,B\!=\!139^\circ,-31^\circ$ in Supergalactic coordinates).
The bulk velocity within $60\hmpc$ is $300-350\kms$ towards
($L,B\!\approx\!166^\circ,-20^\circ$) (\se{de_fluct_bulk}) 
but the flow is not coherent over the whole volume sampled, \eg,
there are regions in front of PP (bottom right)
and behind the GA (far left) where the $XY$
velocity components vanish, \ie, the streaming relative to the LG
is opposite to the bulk flow direction.  The velocity field shows local
convergences and divergences which indicate strong density variations on
scales about twice as large as the smoothing scale.

\begin{figure}[th]
\vspace{11.0truecm}
{\includegraphics{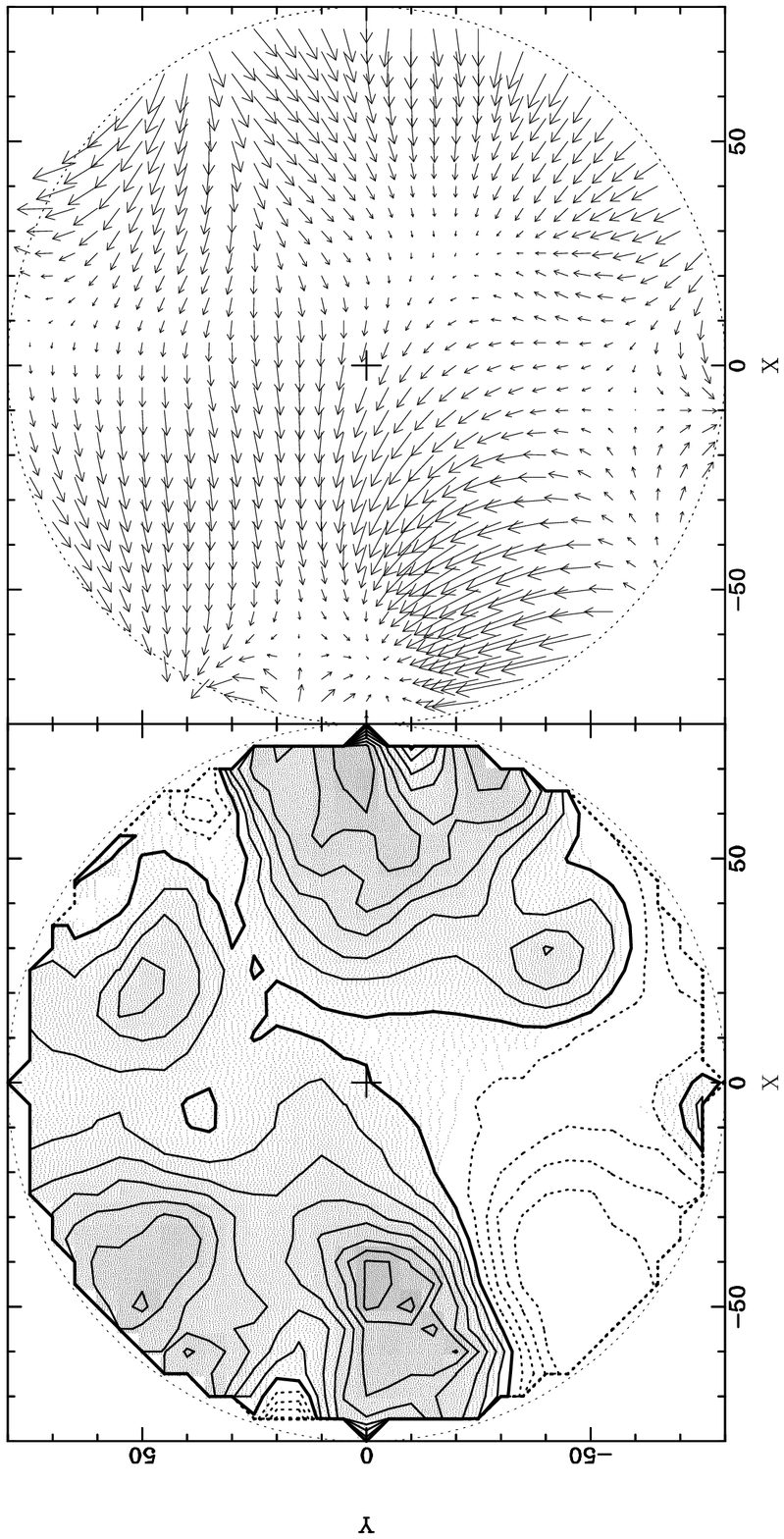}}
{\includegraphics{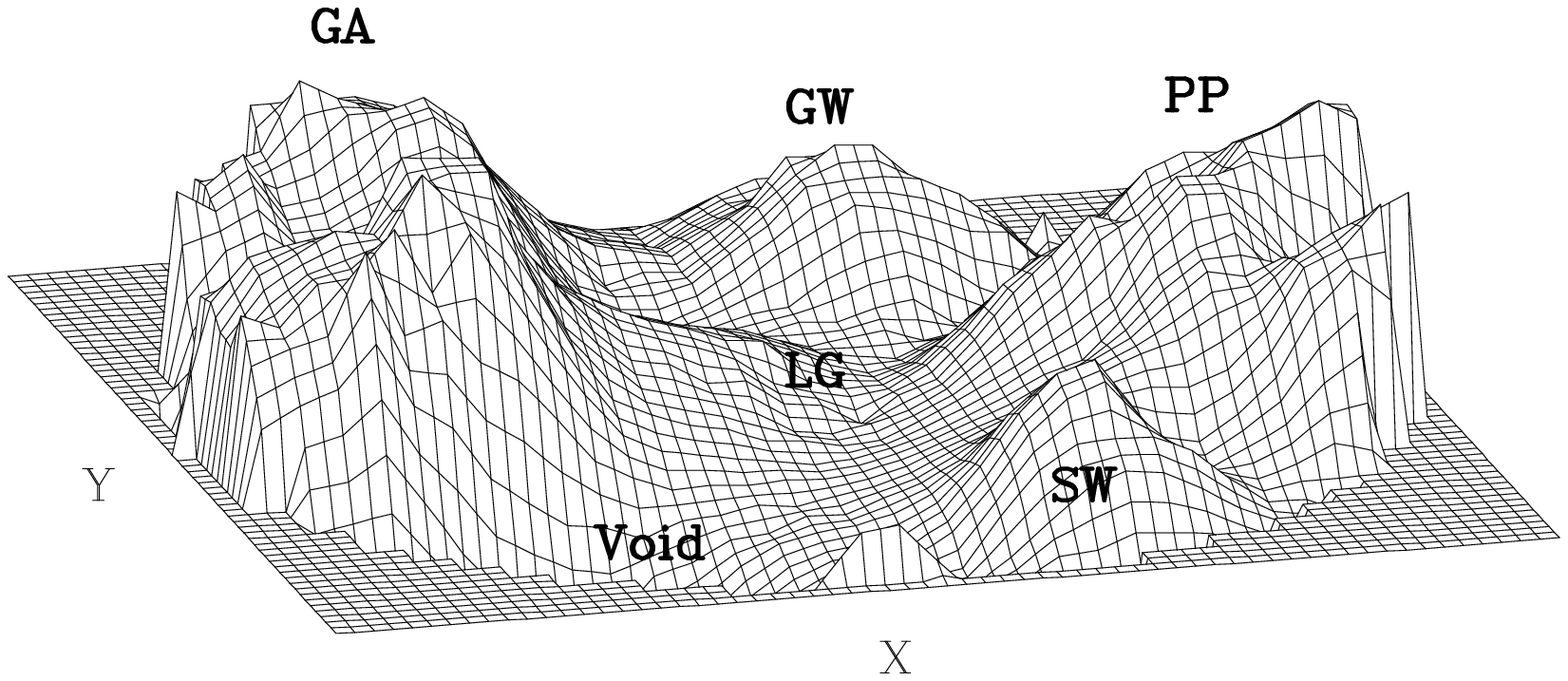}}
\caption{\capt 
POTENT dark-matter maps.
The G12 fluctuation fields of peculiar velocity and {\it mass}-density in the
Supergalactic plane as recovered by POTENT from the Mark III peculiar
velocities.  The vectors are
projections of the 3D velocity field in the CMB frame, dominated by
large derivatives embedded in a coherent bulk flow.
Distances and velocities are in $100\kms$. Contour spacing
is 0.2 in $\delta$, with the heavy contour marking $\delta=0$ and dashed
contours $\delta<0$.
The height in the surface plot is proportional to the total mass density
contrast $\delta$.
The LG is at the center,
GA on the left, PP and the Southern Wall on the right, Coma Great Wall
at the top, and the Sculptor void in between (Dekel \etal 1997).
}
\label{fig:de_pot_map}
\end{figure}

The bottom panel of Fig.~\ref{fig:de_pot_map} shows the POTENT mass
density field in the Supergalactic plane as a landscape plot.
The Great Attractor (with $12\hmpc$ smoothing and $\Omega\!=\!1$)
is a broad density ramp
of maximum height $\delta\!=\!1.4\pm 0.3$ located near the Galactic plane
$Y\!=\!0$ at $X\!\approx\!-40\hmpc$.  The GA extends towards Virgo near
$Y\!\approx\!10$
(the ``Local Supercluster"), towards Pavo-Indus-Telescopium (PIT) across
the Galactic plane to the south ($Y\!<\!0$), and towards the Shapley
concentration behind the GA ($Y\!>\!0, X\!<\!0$). The structure at the top
is related to the ``Great Wall'' of Coma, with
$\delta\approx0.6$.  The Perseus Pisces peak which dominates the right-bottom
is peaked near Perseus with
$\delta\!=\!1.0\pm0.4$.  PP extends towards the southern
galactic hemisphere (Aquarius, Cetus), coinciding with the
``Southern Wall" as seen in redshift surveys.  Underdense regions
separate the GA and PP, extending from bottom-left to top-right.  The
deepest region in the Supergalactic plane, with $\delta\!=\!-0.8\pm0.2$,
roughly coincides with the galaxy-void of Sculptor and is useful in bounding
$\Omega$ (\se{de_omega}).

\section{STATISTICS OF MASS-DENSITY FLUCTUATIONS}
\label{sec:de_fluct}

 Having assumed evolution by \gi, the structure can be traced backward in
time in order
to recover the initial fluctuations and to measure statistics which
characterize them as a random field, \eg, the power spectrum, \pk,
and the probability distribution functions (\pdf).
``Initial" here may refer either to the {\it linear}
regime at $z\!\sim\!10^3$ after the onset of the self-gravitating matter
era, or to the origin of fluctuations in the early universe before
being filtered on sub-horizon scales during the plasma-radiation era.
The spectrum is filtered on scales $\leq\!100\hmpc$ by dark-matter dominated
processes, but its shape on scales $\geq\!10\hmpc$ is not
affected much by the mildly-nonlinear effects (because the faster density
growth in superclusters roughly balances the slower density depletion
in voids at the same wavelength).
The shape of the one-point \pdf, on the other hand, is expected to survive
the plasma era unchanged but it develops strong skewness even in the
mildly-nonlinear regime.  Thus, the present day \pk\ can be used as is to
constrain the origin of fluctuations (on large scales) and the nature of the
\dm\ (on small scales), while the \pdf\ needs to be traced back to the linear
regime first.

The competing scenarios of \lss\ formation are reviewed for example by Primack
(1997).
In summary, if the dark matter (DM) is all baryonic, then by nucleosynthesis
constraints the universe must be of low density,
$\Omega \lsim 0.1$, and a viable model for \lss\ is the Primordial
Isocurvature Baryonic model (\pib) with several free parameters.
With $\Omega\!\sim\!1$, the non-baryonic \dm\
constituents are either ``cold" or ``hot", and the main competing
models are \cdm, \hdm, and \chdm\ --- a 7:3 mixture of the two.
The main difference in the \dm\ effect on \pk\ arises
from free-streaming damping of the ``hot" component of fluctuations on
galactic scales.  Currently popular variants of the standard \cdm\ model
($\Omega\!=\!1$, $n\!=\!1$) include a tilted power
spectrum on large scales ($n\lsim 1$) and a flat, low-$\Omega$ universe with
a non-zero cosmological constant such that $\omm+\oml=1$.

The peculiar velocities of the Mark III catalog enable direct derivations
of the mass power spectrum itself, independent of galaxy biasing,
roughly in the range $10\!-\!100\hmpc$.
The bulk velocity in spheres of radii up to $60\hmpc$ is sensitive
to even larger wavelengths.
In all standard theories, the power spectrum on large scales is expected
to be a power law, $P_k\propto k^n$, with $n$ of order unity. It is expected
to turn around at $k_{peak} \sim 0.065 (\Omega h)^{-1} (\hmpc)^{-1}$,
corresponding to the horizon scale at the epoch of equal energy densities
in matter and radiation. The dark matter type mostly affects the shape
of the filtered spectrum in the ``blue" side of the peak ($k>k_{peak}$).
Once the fluctuation amplitude on very large-scales is fixed by COBE's
measurements of CMB fluctuations,
the bulk velocity is sensitive to $n$ and is insensitive to $\Omega$ or the
DM type.
The steep slope of the CDM-like spectra at $k>k_{peak}$, where it is
best constrained by the data, makes it more sensitive than the bulk
velocity to $\Omega h$.

We first describe the bulk velocity
(\se{de_fluct_bulk}). Then two ways of evaluating \pk:
a model-independent evaluation
from the velocity field via POTENT (\se{de_fluct_pot}),
and a likelihood estimation 
from raw radial peculiar velocities under a prior model
(\se{de_fluct_cobe}).
The \pk\ from the local velocities is then compared to sub-degree angular
power spectrum of \cmb\ fluctuations (\se{de_fluct_cmb}).

\subsection{Bulk Velocity}
\label{sec:de_fluct_bulk}

A simple and robust statistic related to the power spectrum is the bulk
velocity --- the amplitude of the vector average $\vV$ of the
$R_s$-smoothed velocity field $\vv(\vx)$ over a volume defined by a normalized
window function $W_R(\vr)$ (\eg, top-hat) of a characteristic scale $R$,
\begin{equation}
\vV\equiv\int d^3x\, W_R(\vx)\, \vv(\vx) \ ,
\ \ \
\langle V^2 \rangle = {f(\Omega)^2 \over 2 \pi^2} \int_0^\infty dk\, P(k)\,
\widetilde{W}_R^2(\vk) \ .
\label{eq:de_bulk}
\end{equation}
We denote by $V_r$ the bulk velocity in a top-hat sphere of radius
$R=r\hmpc$.
The ensemble variance $\langle V^2 \rangle$ for a model that is characterized
by \pk\ is an integral of \pk\ in which the wavelengths
$\geq\!R$ are emphasized by $\widetilde{W}_R^2(\vk)$, the Fourier
transform of $W_R(\vr)$.
The bulk velocity can be obtained from the observed radial velocities by
minimizing residuals as in Eq.~\ref{eq:de_sumv}.
The first report by Dressler \etal (1987) of $V\!=\!599\pm104$
for ellipticals within $\sim\!60\hmpc$
was interpreted prematurely as being in severe excess of common predictions,
but it quickly became clear that the effective window was much smaller
due to the nonuniform sampling and weighting
(Kaiser 1988).
The sampling-gradient bias can be crudely corrected by volume weighting as in
POTENT (\se{de_rec_pot_twf}), at the expense of larger noise.
Courteau \etal (1993) reported based on an early version
of the Mark III data
$V_{60}=360\pm40$ towards $(L,B)=(162^\circ,-36^\circ)$.
%
Alternatively, $\vV_r$ can be computed from the POTENT $\vv$ field
by simple vector averaging from the grid.

The bulk velocity as a function of $R$,
from several recent sources, is shown in Figure \ref{fig:de_bulk}.
The Mark III POTENT result at $R=50\hmpc$ is
$V_{50}=374\pm85\kms$ towards $(158^\circ,-9^\circ) \pm 10^\circ$.
The $\sim\! 20\%$ error bars are due to distance errors, and
one should consider an additional uncertainty of similar
magnitude due to the non-uniform sampling.
The SFI sample of Sc galaxies yields at the same $R$
a very similar result (contrary to premature rumors),
$V_{50}\approx364\kms$ towards $(172^\circ,-14^\circ)$ (daCosta \etal 1996).
These samples are not large enough for a reliable estimate of $V$
at larger radii.

\begin{figure}[th]
\vspace{6.5truecm}
{\includegraphics{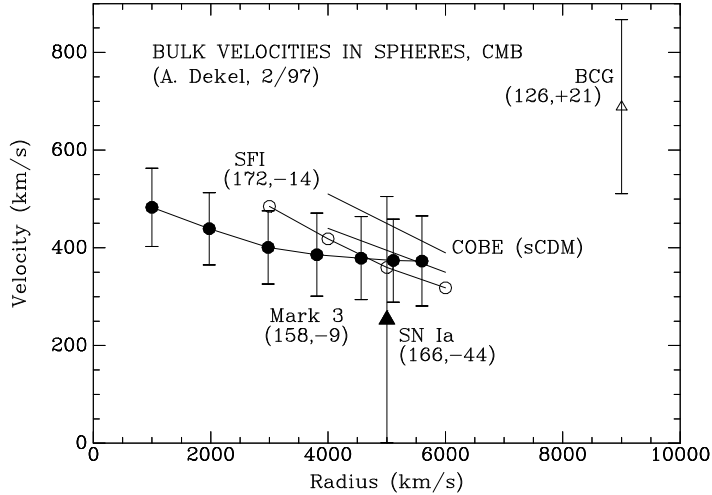}}
\caption{\capt 
Bulk velocity.
The amplitude of the bulk velocity relative to the CMB frame
in top-hat spheres about the LG, as derived from several data sets.
The directions are indicated in supergalactic coordinates $(L,B)$.
The Mark III data and the SFI data
yield consistent results. The new result from Supernovae type Ia
on larger scales is a natural extrapolation. The result from
brightest galaxies in clusters (LP) is discrepant at more than the $2-\sigma$
level.
}
\label{fig:de_bulk}
\end{figure}

Supernovae Type Ia provide more accurate distances, with only $\sim 8\%$
error, and they can be measured at larger distances. 
The current sample of 44 such SNe by Riess \& Kirshner (1997, following
Riess, Press \& Kirshner 1995),
which extends out to $\sim 300\hmpc$,
shows a bulk flow of 
$V=253 \pm 252\kms$ towards $(166^\circ,-44^\circ)$.
The effective radius of this data set for a bulk flow fit
is in fact less than $50\hmpc$ because the data is weighted
inversely by the errors.
The SNe bulk flow is consistent with the results from the Mark III and SFI
galaxy data.  They all make a bulk
of sense within the framework of standard isotropic and homogeneous
cosmology.

The only apparently discrepant result comes from the velocities
measured on a larger scale using brightest cluster galaxies (BCG)
as distance indicators (Lauer \& Postman 1994, LP).
They indicate a large bulk velocity of $V=689\pm178$ towards a very different
direction $\sim(126,21)$. An ongoing effort to measure BCG's in a larger
sample of clusters and distances to clusters based on other distance
indicators will soon tell whether this early result is a $\gsim
2\sigma$ statistical fluke (Watkins \& Feldman 1995; Strauss \etal 1995),
whether the errors were underestimated, or whether something is systematically
different between the BCG distances and the distances measured by other
indicators.

Shown in comparison are the expected \rms bulk velocity in a standard
CDM model ($\Omega=1$, $n=1$) normalized to COBE, for $h=0.5$ (bottom) and
$0.8$ (top) (Sugiyama 1995).
These theoretical curves would not change much if a 20\% hot component
is mixed with the cold dark matter, or if $\Omega$ is lower but still
in the range $0.2-1.0$, as long as $n\approx1$.
The main effect of $\Omega$ and $H_0$ on \pk\ is via $k_{peak}$.
The predicted bulk velocity over $\sim 100\hmpc$
is effectively an integral of \pk\
over $k<k_{peak}$, and is therefore relatively
insensitive to $\Omega$ while it is quite sensitive to $n$.
When compared to a theoretical prediction, the error should also include
cosmic scatter due to the fact that only one sphere has been sampled
from a random field. These errors are typically on the order of the
measurement errors.

The measurements of \cmb\ fluctuations on scales $\leq\!90^\circ$
are independent of the local streaming motions, but
\gi\  predicts an intimate relation between their amplitudes.
The \cmb\ fluctuations are associated with fluctuations in gravitational
potential, velocity and density in the surface of last scattering at
$z\!\sim\!10^3$, while similar fluctuations in our
neighborhood have grown by gravity to produce the
dynamical structure observed locally.
The comparison between the two is therefore a crucial test for \gi.

Before COBE, the local streaming velocities served to predict
the expected level of \cmb\ fluctuations.
The local surveyed region of $\sim\!100\hmpc$ corresponds to a
$\sim\!1^\circ$ patch on the last-scattering surface.
An important effect on scales $\geq\!1^\circ$
is the Sachs-Wolfe effect (1967), where
potential fluctuations $\Delta \phig$ induce temperature
fluctuations via gravitational redshift,
$\delta T/T \!=\! \Delta \phig /(3c^2)$.
Since the velocity potential is proportional
to $\phig$ in the linear and mildly-nonlinear regimes,
$\Delta \phig \sim\!V x$, where $x$ is the
scale over which the bulk velocity is $V$.
Thus $\delta T/T \geq V\, x / (3c^2)$.
A typical bulk velocity of $\sim\!300\kms$ across $\sim\! 100\hmpc$
(\se{de_fluct_bulk})
corresponds to $\delta T/T \!\geq\! 10^{-5}$ at $\sim\! 1^\circ$.
If the fluctuations are roughly scale-invariant ($n=1$),
then $\delta T/T\!\geq\!10^{-5}$ is expected on all scales $>\!1^\circ$.
Bertschinger, Gorski \& Dekel (1990) produced a crude $\delta T/T$ map
of the local region as seen by a hypothetical distant observer, and predicted
$\delta T/T\!\geq\!10^{-5}$ from the local potential well associated with
the GA. An uptodate version of the $\delta T/T$ maps is provided
by Zaroubi \etal (1997b), who added a proper treatment of the acoustic effects
on sub-degree scales for various cosmological models.

Now that \cmb\ fluctuations of $\sim\! 10^{-5}$ have been detected
practically on all the relevant angular scales,
the argument can be reversed:
if one assumes \gi, then the {\it expected}
bulk velocity in the surveyed volume is $\sim\!300\kms$, \ie, the
inferred motions of \se{de_rec} are most likely real.
If, alternatively, one accepts the peculiar velocities
as real for other reasons,
then their consistency with the CMB fluctuations is a relatively
sensitive and robust test of the validity of \gi. This test is unique
in the sense that it addresses the specific fluctuation growth rate
as predicted by \gi\  theory (\se{de_hypo_dd}).
It is robust in the sense that it is quite insensitive to the values of
the cosmological parameters and is independent of the complex issues
involved in the process of galaxy formation.

\subsection {Power Spectrum from the Velocity Field via POTENT}
\label{sec:de_fluct_pot}

One way to compute the power spectrum is via the smoothed mass density field 
as recovered by POTENT (Kolatt \& Dekel 1997). 
The key is to correct the result for systematic deviations from the 
true \pk.
The data suffers from distance errors and sparse, nonuniform sampling,
and they were heavily smoothed.
The \pk\ is computed from within a window of effective radius
$\sim 50\hmpc$, say, where the densities are weighted 
inversely by the squares of the local errors.
The density field is zero-padded in a larger periodic box in order to 
enable an FFT procedure. 
The \pk\ is computed by averaging the amplitudes of the Fourier 
transforms in bins of $k$. This procedure yields an ``observed" \pk, 
which we term $O(k)$.

The systematic errors in the above procedure are then modeled by
$O(k)=M(k) [S(k) + N(k)]$,
where $S(k)$ is the true signal \pk, $N(k)$ is the noise,
and $M(k)$ represents the effects of sampling, smoothing, applying a
window etc.
The correction functions $M(k)$ and $N(k)$ can be derived from Monte Carlo
mock catalogs (\se{de_rec_pot_mock}).
The factor $M(k)$ is derived first by
$M^{-1}=S/\langle O \rangle_{no-noise}$, where $S$ here is the
known power spectrum built into the simulations, and the averaging is over
mock catalogs not perturbed by noise. Then $N(k)$ is computed by
$N=M^{-1} \langle O \rangle_{noise} - S$, where the averaging is over
noisy mock catalogs.

\begin{figure}[th]
\vspace{6.5truecm}
{\includegraphics{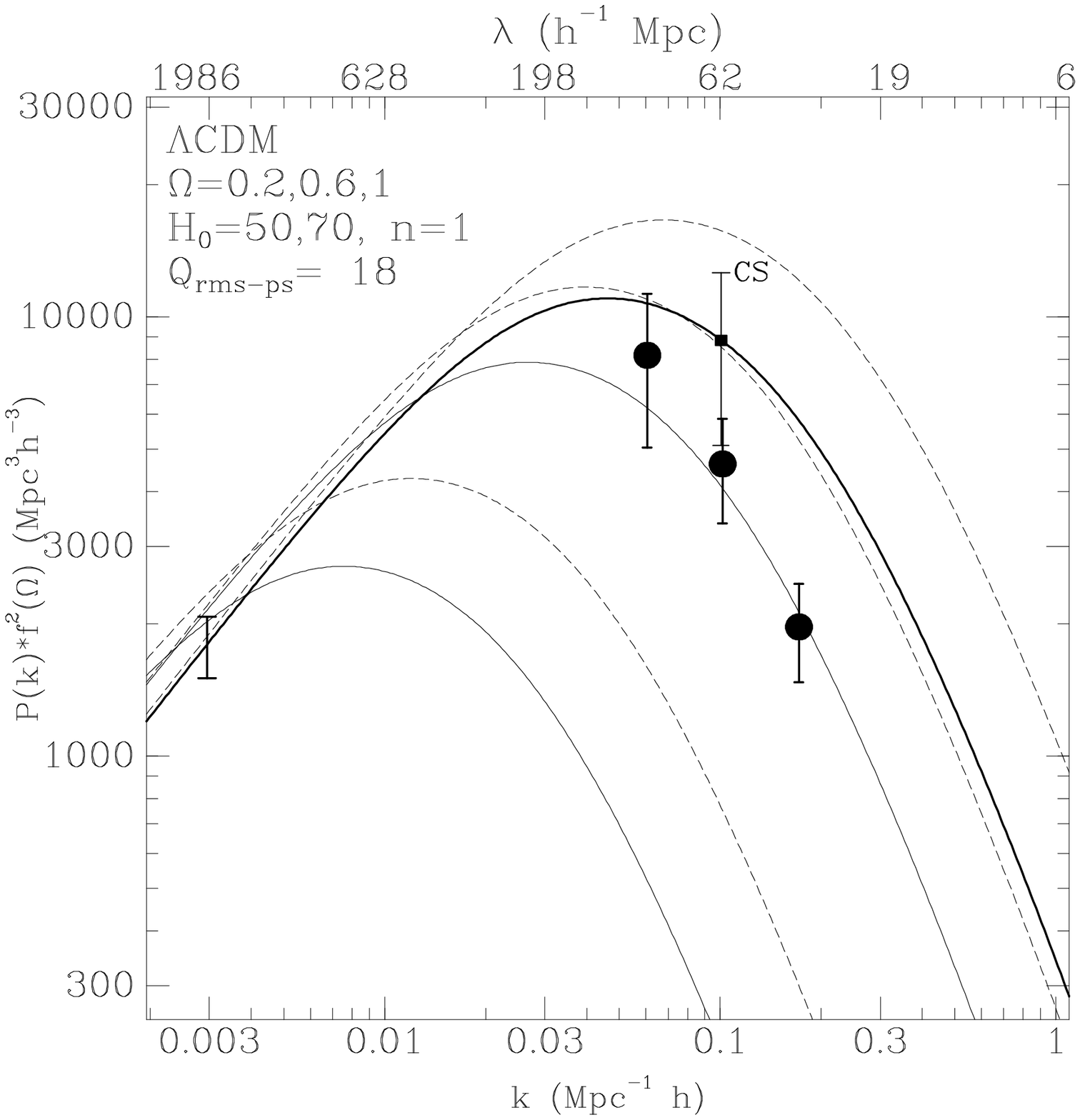}}
{\includegraphics{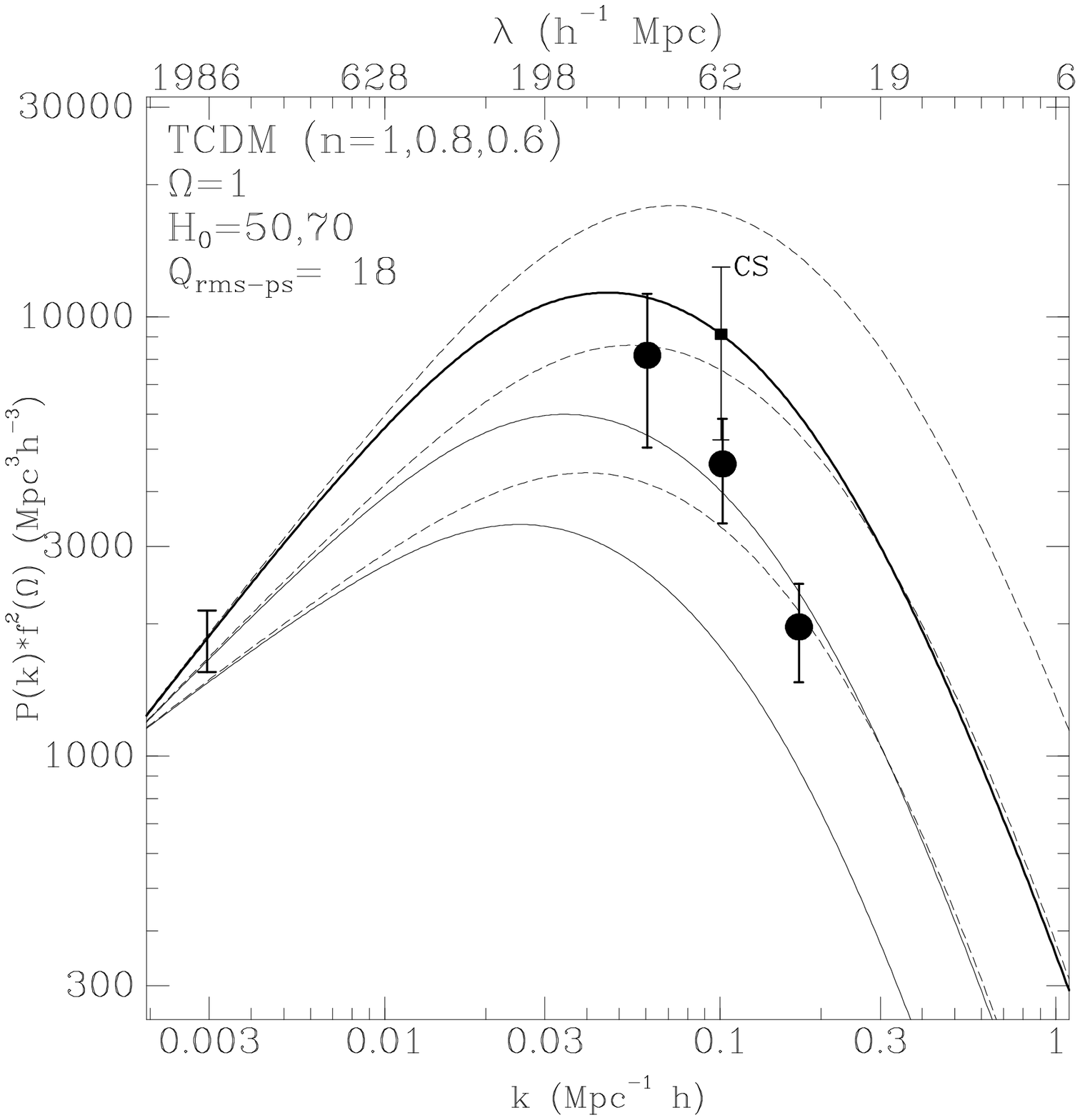}}
\caption{\capt 
The mass power spectrum [$\times f(\Omega)^2$]
from POTENT Mark III velocities (filled symbols),
with $1\sigma$ random errors.
The curves are COBE-normalized theoretical predictions for
flat CDM with $h=0.5$ (solid) and $h=0.7$ (dashed).
Left: $\Lambda$-CDM with $n=1$ and $\Omega$ growing upwards.
Right: tilted-CDM with $\Omega=1$ and $n$ growing upwards.
Typical cosmic scatter (CS) for $\Omega=1$ and $n=1$ is indicated
(Kolatt \& Dekel 1997).
}
\label{fig:de_spec}
\end{figure}

Equipped with the correction functions $M(k)$ and $N(k)$,
the \pk\ observed from the real universe, $O(k)$, is corrected
to yield the true \pk\ by
$S(k) = M(k)^{-1} O(k) - N(k)$.
The recovered mass-density \pk\ is shown in Figure \ref{fig:de_spec}
in three thick logarithmic bins
covering the range $0.04 \leq k \leq 0.2\, (\hmpc)^{-1}$, within which
the results are reliable. The robust result is
$P(k) f(\Omega)^2 =(4.6 \pm 1.4) \times 10^3 (\hmpc)^3$ at $k=0.1
(\hmpc)^{-1}$
(using the convention where the Fourier transform is defined with
no $2\pi$ factors in it's coefficient).
The logarithmic slope at $k=0.1$ is $-1.45\! \pm \! 0.5$.
This translates to $\sigma_8 \,\Omega^{0.6} \simeq 0.7-0.8$,
depending on where the peak in \pk\ is (see \se{de_fluct_cobe}).

The observed \pk\ is compared in the figure to the linear predictions of
a family of Inflation-motivated flat CDM models ($\omm+\oml=1$),
normalized by the 4-year COBE data,
with the Hubble constant fixed at $h=0.5$ or $h=0.7$.  For $n=1$,
maximum likelihood is obtained at
$\Omega \simeq 0.7 h_{50}^{-1.3} \pm 0.1$.  
For $\Omega=1$, assuming no tensor fluctuations,
the linear power index is 
$n \simeq 0.75 h_{50}^{-0.8} \pm 0.1$. 

\subsection{Power Spectrum from Velocities for COBE-CDM Models}
\label{sec:de_fluct_cobe}

The power spectrum, 
in a parametric form including $\Omega$, $h$ and $n$ among the
parameters, has alternatively been determined from the velocity data
via a Baysian likelihood analysis (Zaroubi \etal 1997a; 
see also Kaiser \& Jaffe 1995).
According to Bayse, the probability of the model parameters ($m$) given the
data ($d$), which is the function one wishes to maximize, can be expressed as
$
P(m|d) = P(d|m)\, P(m)/P(d)
$.
The probability $P(d)$ serves here as a normalization constant.
Without any external constraints on the model parameters,
one assumes that $P(m)$ is a constant in a given range.
The remaining task is to maximize the likelihood ${\cal L}=P(d|m)$
as a function of the model parameters. This function can be 
written down explicitly.

Under the assumption that the velocities and the errors are both Gaussian
random fields with no mutual correlations, the likelihood can be written
as
$
{\cal L} = (2 \pi |D|)^{-1/2} \exp (-d_i D_{ij}^{-1} d_j /2)
$,
where $d_i$ are the data at points $i=1,...,N$,
and $D_{ij}$ is the covariance matrix, which can be split into
covariance of signal ($s$) and covariance of noise ($n$),
$
D_{ij} \equiv \langle d_i d_j \rangle
=\langle s_i s_j \rangle  + \langle n_i n_j \rangle
$.
If the errors are uncorrelated, the noise matrix is diagonal.
The signal matrix is computed from the model \pk as
a function of the model parameters.

Zaroubi \etal (1997a) used a parametric model for the PS of the general form
$
P_k = A\, k^n\, T(\Gamma_i;k)
$,
where $T(k)$ is a small-scale filter of an assumed shape characterized
by free parameters $\Gamma_i$,
$k^n$ is the initial \pk\ which is still valid on large scales today,
and $A$ is a normalization factor.
The normalization can either be determined by COBE's data
(for given $\Omega$, $\Lambda$,
$h$, $n$ and tensor/scalar fluctuations), or be left as a free parameter
to be fixed by the velocity data alone.
The filter $T(k)$ can either be taken from a specific physical model
(\eg, CDM, where $\Gamma = \Omega h$), or be an
arbitrary function with enough flexibility to fit the data.

The robust result for all the models is a relatively high amplitude, with
$P(k) f(\Omega)^2 = (4.8\pm1.5)\times 10^3 (\hmpc)^3$ at $k=0.1(\hmpc)^{-1}$.
An extrapolation to smaller scales using the different CDM models yields 
$\sigma_8 \Omega^{0.6} = 0.88 \pm 0.15$
(for the dispersion in top-hat spheres of radius $8\hmpc$).

Within the general family of CDM models, allowing for a cosmological constant
in a flat universe and a tilt in the spectrum, the parameters are confined
by a 90\% likelihood contour of the sort
$\Omega\, {h_{50}}^\mu\, n^\nu = 0.8 \pm 0.2$,
where $\mu = 1.3$ and $\nu = 3.4,\ 2.0$ for models with and without
tensor fluctuations respectively.
Figure~\ref{fig:de_likelihood} displays the likelihood map in the $\Omega-n$
plane for these models.
For open CDM the powers are $\mu = 0.95$ and $\nu = 1.4$
(no tensor fluctuations). 
A $\Gamma$-shape model free of COBE normalization yields
only a weak constraint: $\Gamma = 0.4\pm 0.2$ (where $\Gamma$ is not
necessarily $\Omega h$).

\begin{figure}[th]
\vspace{6.0truecm}
{\includegraphics{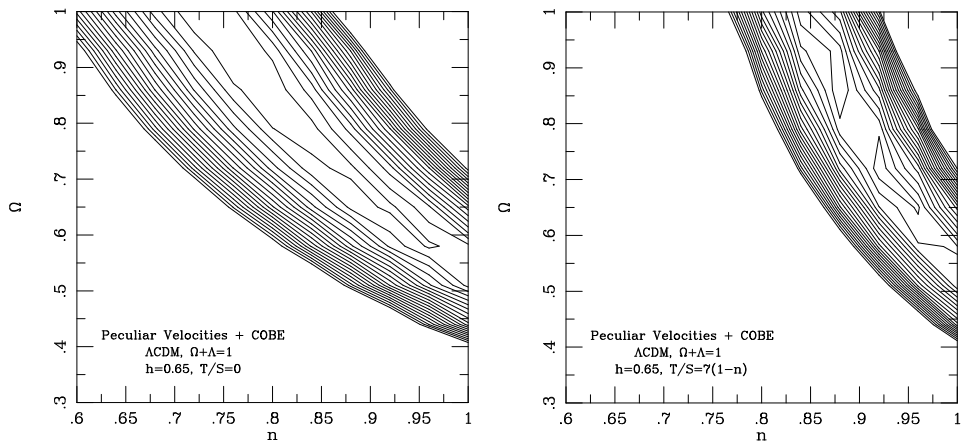}}
\caption{\capt 
Likelihood contour maps in the $\Omega-n$ plane for flat CDM with
and without tensor fluctuations and $h=0.65$.
Contour spacing is unity in log-likelihood. 
Under the assumption of a $\chi^2$ distribution, 
the two-dimensional 90 percentile corresponds to 2.3 contours,
and the one-dimensional 90 percentile corresponds to 1.35 contours
(Zaroubi \etal 1997a).
}
\label{fig:de_likelihood}
\end{figure}

Both $\Omega$ and $n$ obtained by the likelihood analysis
from the raw peculiar velocities tend to be slightly higher ($\sim\! 20\%$)
than their estimates based on
the \pk\ recovered from the POTENT output.
This difference may arise from the different relative weighting
assigned to the different wavelengths in the two analyses.
The difference between the results obtained in the two different ways
is on the order of the errors in each analysis and the cosmic
scatter.
Very similar estimates of \pk\ are obtained from a preliminary analysis
of the SFI sample of Sc galaxies (in preparation). 

In summary:
The ``standard" CDM model is marginally rejected at the $\sim\!2\sigma$
level, while each of the following modifications lead to a good fit to the
peculiar velocities and large-scale CMB data:
$n\lsim 1$, $\Omega_\nu\sim 0.3$, or $\Omega \lsim 1$.
The strong implication on the dark matter issue is that
values of $\Omega$ as low as $\sim\! 0.2$ are ruled out with high
confidence (independent of $\Lambda$), leaving, in particular,
no room for the baryonic PIB model.

\subsection {Peculiar Velocities vs Small-scale CMB Fluctuations} 
\label{sec:de_fluct_cmb}

Sub-degree angular scales at the last scattering surface correspond 
to the $\leq 100\hmpc$ comoving scales explored by peculiar velocities today. 
Thus, under the assumption that the local neighborhood is typical,
the power spectrum on these scales is simultaneously constrained 
by the mass-density fluctuations in our cosmological neighborhood
and by the CMB fluctuations.

The sub-degree CMB fluctuations are now being explored
by many balloon-born experiments, and in less than a decade we expect
accurate results from the CMB satellites MAP and Planck.
These measurements will eventually allow
a simultaneous likelihood analysis of the two kinds of data.  
At this point, however, although there are already preliminary
detections of the first acoustic peak in the angular power spectrum,
the uncertainties are still large. Any current comparison is therefore
limited to the semi-quantitative level.
The range of parameters permitted by the peculiar velocity data
for power spectra of the CDM family (Zaroubi \etal 1997a)
can be translated to a range of angular power spectra, $C_l$. 
This range is plotted against current observations in 
Figure \ref{fig:de_cmb_cl}. 

\begin{figure}[th]
\vspace{6.0truecm}
{\includegraphics{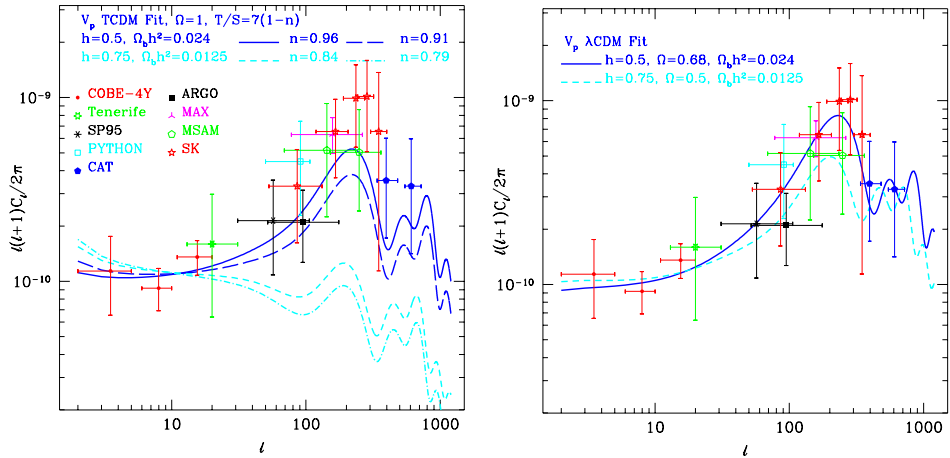}}
\caption{\capt 
Angular power spectrum in the CMB. The current data (symbols)
are compared to certain
CDM models that fit the peculiar velocity data at 90\% likelihood.
The extreme models shown are either with low $h$ and high $\omb h^2$
or vice versa.
Left: tilted CDM with $\Omega=1$, including tensor fluctuations.
Right: Flat CDM with $n=1$ (Zaroubi \etal 1997b).
xxxxxx
}
\label{fig:de_cmb_cl}
\end{figure}

The immediate conclusions from a visual inspection of the figure are that
a wide range of CDM models can simultaneously obey the two data sets,
but there is a subset of models that fits the velocities well but seems
to fail to produce a high enough acoustic peak in the CMB spectrum.

The acoustic peak in the CMB is sensitive to $\omb$ and
the observations prefer a high baryon content,
$\omb h^2\sim 0.025$ (similar to the value measured by Tytler \etal 1996;
Burles \& Tytler 1996),
while the peculiar velocities have little to add because \pk\ is hardly
affected by $\omb$.

The power index $n$ is important in both cases; the peculiar velocities
allow values of $n$ significantly lower than unity, but the current CMB data 
seem not to tolerate values of $n$ below 0.9 or so.  

The peculiar velocity data prefers $\Omega \geq 0.4$,
and the location of the first acoustic peak in the sub-degree
CMB data indicates in agreement a high value of $\omm+\oml$.

\section{DIRECT MEASUREMENTS OF \pmb{$\Omega$}\ FROM PECULIAR VELOCITIES}
\label{sec:de_omega}
 
 Assuming that the inferred motions are real and generated by \gi, they
can be used to estimate $\Omega$ in several different ways.
Most of the evidence from virialized systems on scales $\leq 10\hmpc$
suggest a low mean density of $\Omega\!\sim\!0.2$
(see Dekel, Burstein \& White 1997).
The spatial {\it variations} of the large-scale velocity field provide
ways to measure the mass density in a larger volume that may be
closer to a ``fair" sample.  One family of such methods is based on comparing
the dynamical fields derived from velocities to the fields
derived from galaxy redshifts (\se{de_beta}).  These methods can be
applied in the linear regime but they always rely on the assumed biasing
relation between galaxies and mass often parameterized by $b$, so they
actually provide an estimate of $\beta\!\equiv\! f(\Omega)/b$.
Another family of methods measures
$\beta$ from redshift surveys alone, based on
$z$-space deviations from isotropy (see Strauss 1997).
In the present section, we focus first on
methods that rely on non-linear effects in the peculiar
velocity data {\it alone}, and
they thus provide estimates of $\Omega$ independent of galaxy density biasing.
These methods are based on the assumption 
that the initial fluctuations were Gaussian.

\subsection{Divergence in Voids}
\label{sec:de_omega_void}

 A diverging flow in an extended low-density region can provide a robust
dynamical lower bound on $\Omega$, based on the fact that large gravitating
outflows are {\it not} expected in a low-$\Omega$ universe 
(Dekel \& Rees 1994).
In practice, for any assumed value of $\Omega$,
the partial derivatives of the smoothed observed velocity field 
are used to infer a non-linear approximation for the mass density via the
approximation $\delta_c$ (\ref{eq:de_delc}). A key point is that 
this approximation is typically an overestimate, $\delta_c\!>\!\delta$
(when the true value of $\Omega$ is used). 
For fluctuations that started Gaussian, the probability that $\delta_c$ 
is an overestimate, in the range $\delta < -0.5$, is well over 99\%.
Analogously to $\delta_0\! \approx\! -\Omega^{-0.6} \divv$,
the $\delta_c$ inferred from a given diverging
velocity field is more negative when a smaller $\Omega$ is assumed,
so for a small enough $\Omega$ one may obtain $\delta<-1$ in certain
void regions. Such values of $\delta$ are forbidden because mass is
never negative, so this provides a lower bound on $\Omega$.
 
The inferred $\delta_c$ field, smoothed at $12\hmpc$, and the associated
error field $\sigma_\delta$, were derived by POTENT at grid points from
the observed radial velocities of Mark III. Focusing on the deepest density
wells, the input $\Omega$ was lowered until $\delta_c$ became
significantly smaller than $-1$.
The most promising ``test case" provided by the Mark III data is
a broad diverging region centered near the supergalactic plane at the
vicinity of $(X,Y)\!=\!(-25,-40)$ in $\hmpc$ ---
the ``Sculptor void" of galaxies
between the GA and the ``Southern Wall" extension of PP
(Figure~\ref{fig:de_void}, compare to Fig.~\ref{fig:de_pot_map}).
 
\begin{figure}[th]
\vspace{6.2truecm}
{\includegraphics{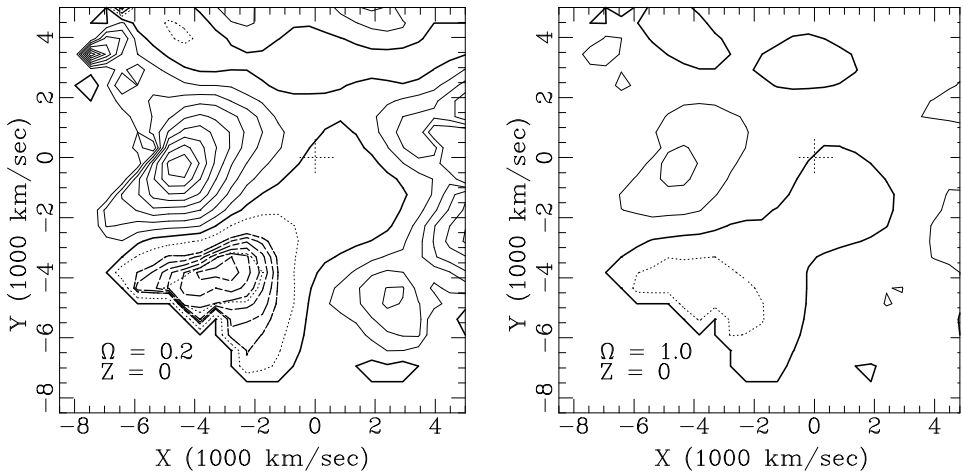}}
\caption{\capt 
Maps of $\delta_c$ inferred from the observed velocities near the
Sculptor void in the Supergalactic plane, for two values of $\Omega$.
The LG is marked by '$+$' and the void is confined by the Pavo part of
the GA (left) and the Aquarius extension of PP (right).  Contour spacing
is 0.5, with $\delta_c=0$ heavy, $\delta_c>0$ solid, and $\delta_c<0$
dotted.  The heavy-dashed contours mark the illegitimate downward
deviation of $\delta_c$ below $-1$ in units of $\sigma_\delta$, starting
from zero (\ie, $\delta_c=-1$) and decreasing with spacing $-0.5\sigma_\delta$.
The value $\Omega=0.2$ is ruled out at the $2.9\sigma$ level (Dekel \&
Rees 1994).
}
\label{fig:de_void}
\end{figure}
 
Values of $\Omega\!\approx\!1$ are perfectly consistent with the data,
but $\delta_c$ becomes smaller than $-1$ already for $\Omega\!=\!0.6$.
The values $\Omega\!=\!0.3$ and $0.2$ are ruled out at the $2.4\sigma$ and
$2.9\sigma$ levels in terms of the random error $\sigma_\delta$.
 
This result is still to be improved. The systematic
errors have been partially corrected in POTENT, but a
more specific investigation of the biases affecting the smoothed
velocity field in density wells is still in progress.
For the method to be effective one needs to find a void that is (a)
bigger than the correlation length for its vicinity to represent the
universal $\Omega$, (b) deep enough for the lower bound to be tight, (c)
nearby enough for the distance errors to be small, and (d) properly
sampled to trace the velocity field in its vicinity.

Note that this method does not require that the void be spherical
or of any other particular shape, and is independent of galaxy density 
biasing. Another pro is that there is no much cosmic scatter --- 
one deep and properly sampled void is enough for
a meaningful constraint. The main limitation is the poor (and perhaps
biased) sampling of the velocity field in the vicinity of a void.

\subsection{Deviations from Gaussian PDF}
\label{sec:de_omega_pdf}

Assuming that the initial fluctuations are a Gaussian random
field, the one-point probability distribution function (PDF)
of smoothed density develops a characteristic skewness due to non-linear
effects early in the non-linear regime (\eg, Kofman \etal 1994).
The skewness of $\delta$ is given according to
second-order perturbation theory by
\begin{equation}
\langle \delta^3\rangle / \langle \delta^2 \rangle^2 
  \approx (34/7\! -3\! -n)\ ,
\label{eq:de_skewd}
\end{equation}
with $n$ the effective power index of the power spectrum near the
(top-hat) smoothing scale (Bouchet \etal 1992).
Since this ratio of moments for $\delta$ is practically independent of
$\Omega$, and since $\divv\!\sim\!-f\delta$, the corresponding ratio
for $\divv$ must strongly depend on $\Omega$, and indeed in second-order
it is (Bernardeau \etal 1995)
\begin{equation}
T_3\equiv
\langle (\divv)^3 \rangle / \langle (\divv)^2 \rangle^2 \approx
-f(\Omega)^{-1} (26/7 -3 -n) \ .
\label{eq:de_skewv}
\end{equation}
Using N-body simulations and $12\hmpc$ smoothing one indeed finds
$T_3\!=\!-1.8\pm0.7$ for $\Omega\!=\!1$ and $T_3\!=\!-4.1\pm1.3$ for
$\Omega\!=\!0.3$,
where the error is the cosmic scatter for a sphere of radius
$40\hmpc$ in a \cdm\ universe ($H_0=75$, $b=1$).  An estimate
of $T_3$ in the current POTENT velocity field within $40\hmpc$
is $-1.1\pm 0.8$, where the
errors this time represent distance errors.  With the two errors added in
quadrature, $\Omega\!=\!0.3$ is rejected at the $\sim\!2\sigma$ level
(somewhat sensitive to the assumed \pk).

\begin{figure}[th]
\vspace{6.3 truecm}
{\includegraphics{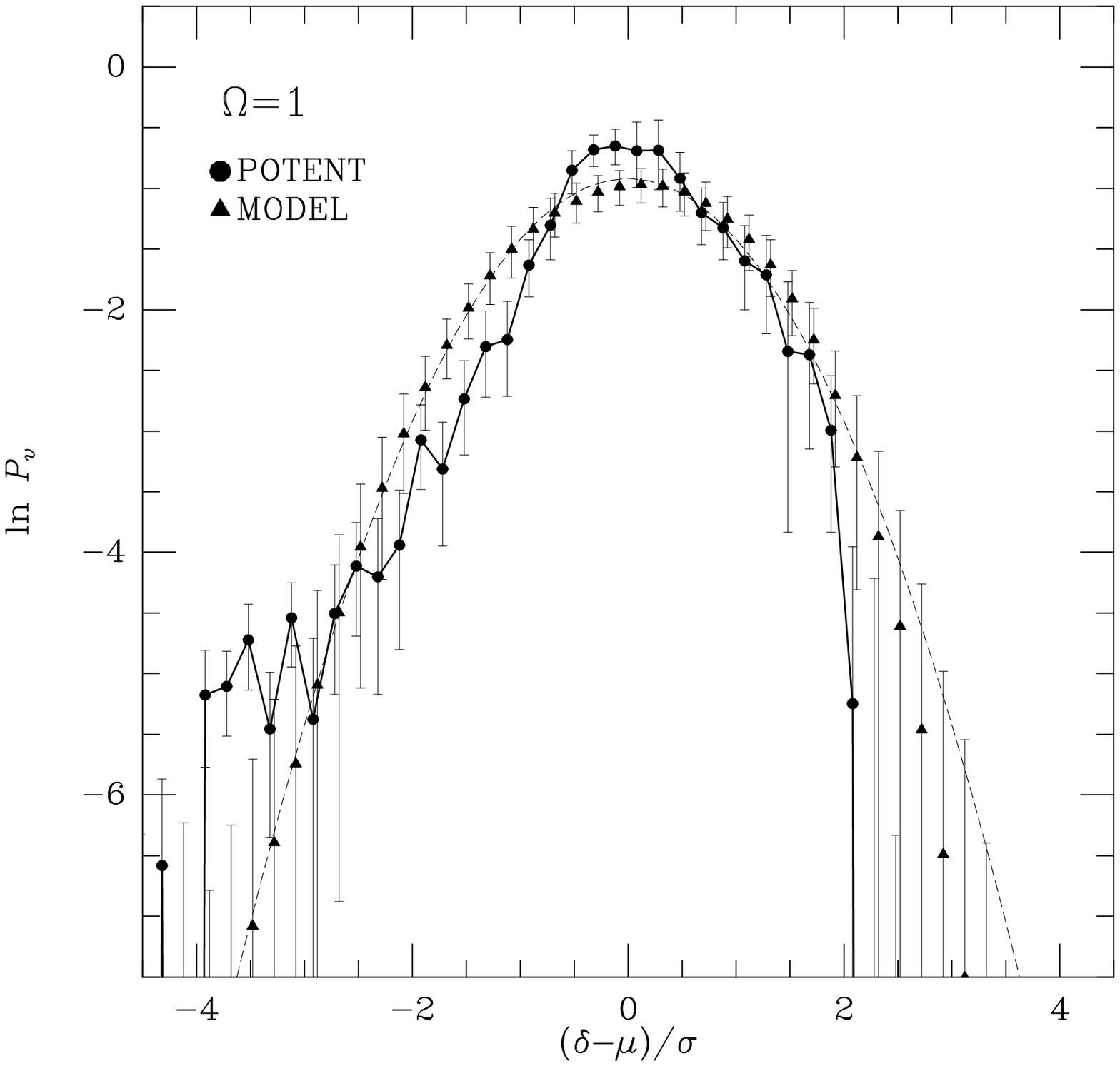}}
{\includegraphics{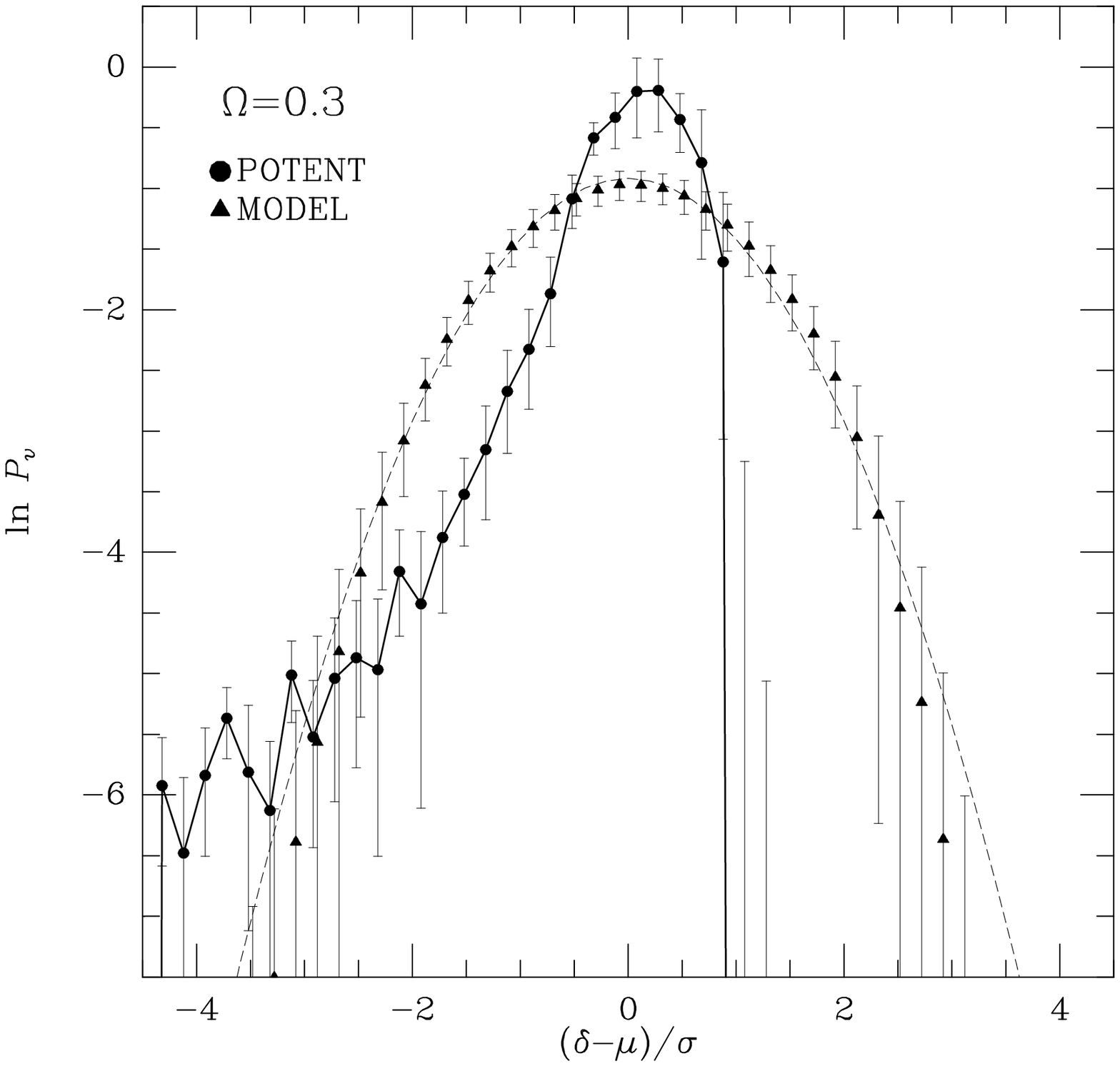}}
\caption{\capt 
$\Omega$ from \ipdf.
The density \ipdf\ recovered from the G2 POTENT peculiar velocity field
(solid), compared to a normal distribution (short dash)
and with the \ipdf\ recovered from the velocity field of Gaussian \cdm\
simulations (triangles).
The assumed $\Omega$ is 1.0 (left) or 0.3 (right).
The simulations are of $\Omega_0=\Omega$ accordingly
(Nusser \& Dekel 1993).
}
\label{fig:de_zpom}
\end{figure}
 
 Since the present PDF contains only part of the information stored in the data
and is in some cases
not that sensitive to the initial PDF (IPDF), a more powerful bound
can be obtained by using the detailed present velocity field $\vv(\vx)$
to recover the IPDF,
and using the latter to constrain $\Omega$ by measuring it's $\Omega$-dependent
deviation from the assumed normal distribution (Nusser \& Dekel 1993).
The necessary ``time machine" is provided by the Eulerian interpretation of
the Zel'dovich approximation (Nusser \& Dekel 1992).

The velocity out of POTENT Mark
II, within a conservatively selected volume, was fed into the \ipdf\
recovery procedure with $\Omega$ either 1 or 0.3, and the errors due to
distance errors and cosmic scatter were estimated.  
Figure \ref{fig:de_zpom} shows the recovered \ipdf's.
The \ipdf\ recovered
for $\Omega=1$ is marginally consistent with Gaussian, while the
one recovered for $\Omega=0.3$ shows significant deviations.  The
largest deviation, bin by bin in the \ipdf, is $\lsim\!2\sigma$ for
$\Omega\!=\!1$ and $>\!4\sigma$ for $\Omega=0.3$, and a similar rejection
of $\Omega=0.3$
is obtained with a $\chi^2$-type statistic.  The skewness and kurtosis
are poorly determined because of noisy tails but the replacements 
$\langle x\vert x\vert\rangle$ and 
$\langle \vert x\vert\rangle$ allow a rejection of
$\Omega=0.3$ at the $(5-6)\sigma$ levels.

The main advantage of the methods based on the \pdf\ is their 
insensitivity to galaxy density biasing.
The main weakness is the need for a ``fair" sample; the cosmic 
scatter is large due to the large smoothing scale within the limited
volume.

\section {GALAXY DENSITY VS. VELOCITIES: \pmb{$\Omega$}\ AND BIASING}
\label{sec:de_beta}

\subsection{Galaxies vs Mass: Fit of GI and Linear Biasing Model}
\label{sec:de_hypo_dd}

\begin{figure}[th]
\vspace{11.4truecm}
{\includegraphics{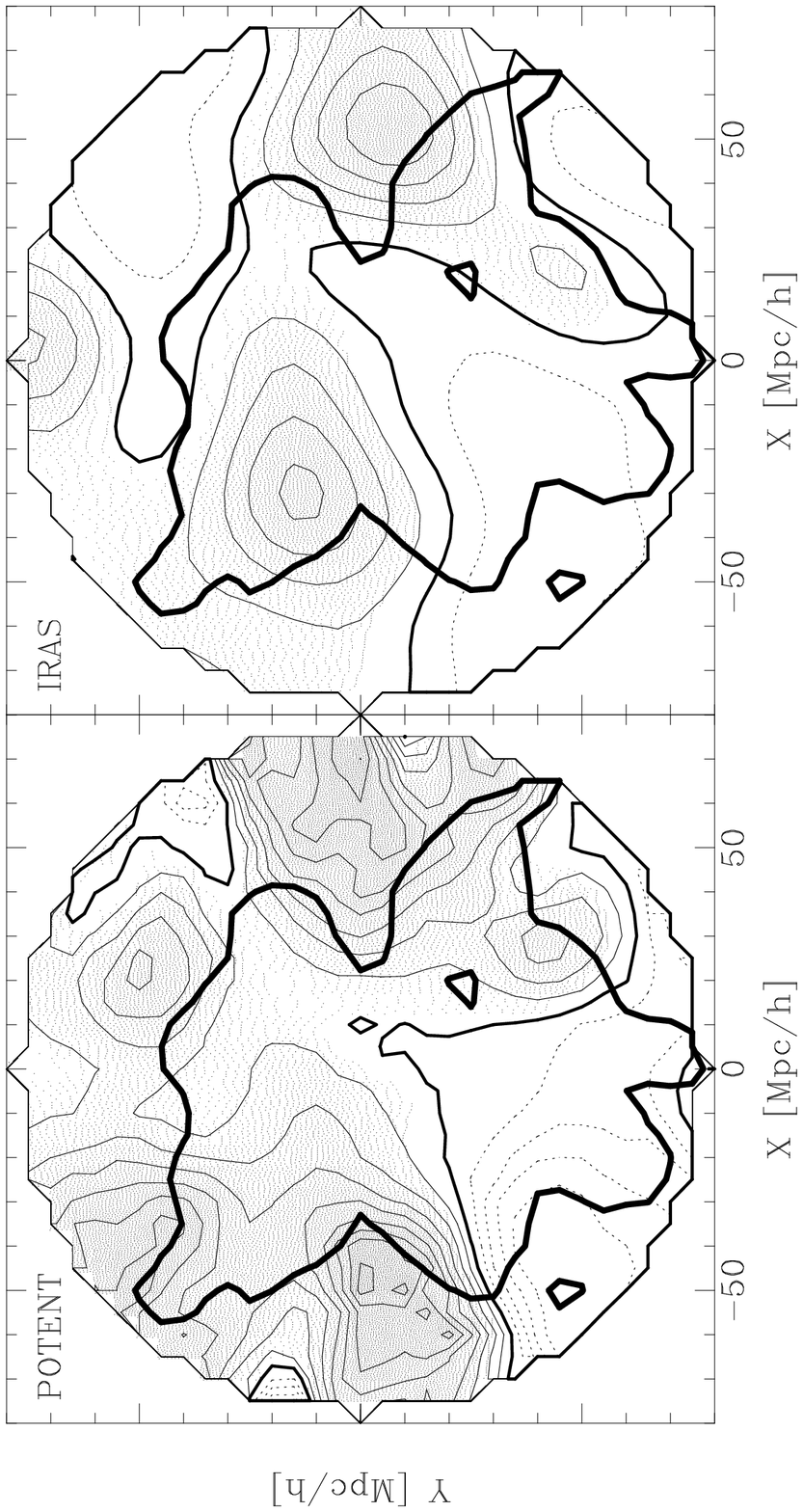}}
{\includegraphics{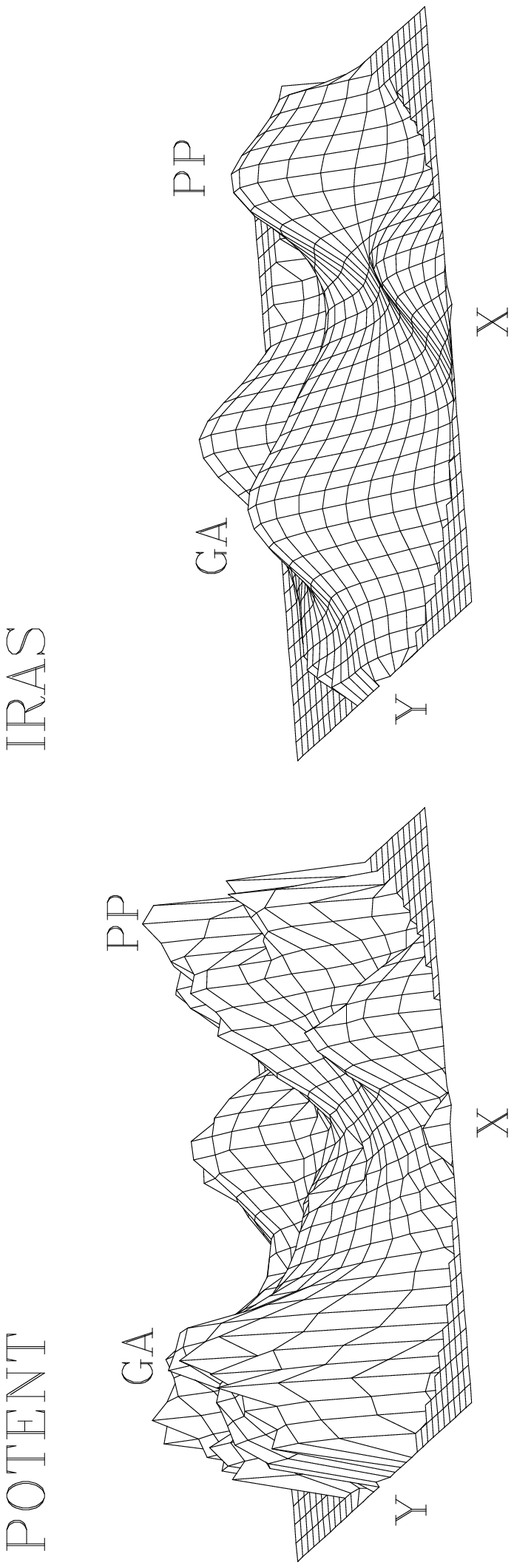}}
\caption{\capt 
Mass versus galaxies.
POTENT mass ($\Omega=1$) versus IRAS galaxy density fields in the Supergalactic
plane, both smoothed G12. Contour spacing is $0.2$.
The heavy contour marks the boundary of the comparison volume of
effective radius $46\hmpc$.
The height in the surface plot is proportional to $\delta$.
The LG is at the center, GA on the left, PP on the right,
and the Sculptor void in between
(Sigad \etal 1997).
}
\label{fig:de_potiras_map}
\end{figure}

 The theory of \gi\ combined with the assumption of linear biasing for
galaxies predict a correlation between the dynamical density field
and the galaxy density field, which can be addressed
quantitatively based on the mock catalogs and the
estimated errors in the two data sets.
Figure \ref{fig:de_potiras_map} 
compares density maps in the Supergalactic plane for IRAS 1.2
Jy galaxies ($\delg$) and POTENT Mark III mass ($\delta$),
both G12 smoothed.
The general correlation is evident --- the GA, PP, Coma
and the voids
all exist both as dynamical entities and as structures of galaxies.
To evaluate goodness of fit,
Figure \ref{fig:de_potiras_chi2}
shows the statistic $\chi^2= N^{-1} \sum^N (\delg - b\delta)^2/\sigma^2$
as computed from the data
in comparison with its distribution over pairs of Mark III and IRAS 1.2 Jy
mock catalogs. The fact that the data lies near the center of this
distribution indicates that the two data sets are consistent
with being noisy versions of an underlying fluctuation field
and that the data
are in agreement with the hypotheses of \gi\  plus linear biasing
(Dekel \etal 1993; Sigad \etal 1997; more in \se{de_beta_dd}).

\begin{figure}[th]
\vspace{7truecm}
{\includegraphics{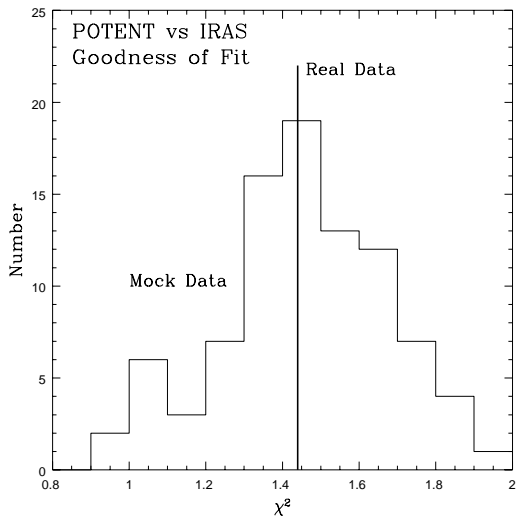}}
\caption{\capt 
Goodness of fit in the comparison of G12 density fields of POTENT mass
and IRAS galaxies, as tested by a $\chi^2$ statistic that is an error
weighted sum of differences between the two fields (Sigad \etal 1997).
}
\label{fig:de_potiras_chi2}
\end{figure}

 What is it exactly that one can learn from the observed $\vv-\delg$
correlation (Babul \etal 1994)?
First, it argues that the velocities are real because it is hard to
invoke any other reasonable way to
make the galaxy distribution and the TF measurements agree so well.
On the other hand, although it is true that gravity is the only
long-range force that could attract galaxies to stream toward density
concentrations, the fact that a $\vv-\delg$ correlation is predicted by \gi\
plus linear biasing does not necessarily mean that the observation
can serve as a sensitive test for either.  Recall that converging
(or diverging) flows tend to generate overdensities (or underdensities)
simply as a result of mass conservation, independent of the source of
the motions.

 Let us assume for a moment that galaxies trace mass, \ie, that the linearized
continuity equation, $\dot{\delta}\!=\!-\divv$, is valid for the galaxies
as well.  The observed correlation (in the linear approximation)
is then $\delta\!\propto\!-\divv$, and
together they imply that $\dot{\delta}\! \propto\! \delta$, or equivalently
that $\divv$ is proportional to its time average.  This property is not
exclusive to \gi; one can construct counterexamples
where the velocities are produced by a non-\gi\ impulse.

Even irrotationality does not follow from $\delta\!\propto\!-\divv$; it
has to be adopted based on theoretical arguments in order to enable
reconstruction from radial velocities or from observed densities.
Once continuity and irrotationality are assumed, the observed
$\delta\!\propto\!-\divv$ implies a system of equations which is
identical in all its {\it spatial} properties to the equations of \gi,
but can differ in the constants of proportionality and their
temporal behavior.  It is therefore impossible to distinguish between \gi\  and
a non-\gi\  model which obeys continuity plus irrotationality
based only on snapshots of
{\it present-day} linear fluctuation fields.  This makes the
relation between \cmb\ fluctuations and velocities
an especially important test for \gi.

On the other hand, the fact that the constant of proportionality in
$\delta\!\propto\!-\divv$ is indeed the same everywhere is
a non-trivial requirement from a non-\gi\  model.  For example,
a version of the explosion scenario (Ostriker \& Cowie 1981; Ikeuchi 1981),
which tested successfully both for irrotationality and
$\vv\!-\!\delta$ correlation,
requires special synchronization among the explosions (Babul \etal 1994).

So, what is the $\vv\!-\!\delg$ relation good for?
While it's sensitivity to \gi\ is only partial, this relation turns out to be
quite sensitive to the validity of a {\it continuity}-like relation
for the {\it galaxies}. When the latter is strongly
violated all bets are off for the $\vv\!-\!\delg$ relation.
A non-linear biasing scheme
would make continuity invalid for the galaxies, which
would ruin the $\vv\!-\!\delg$ relation even if \gi\  is valid.
The observed correlation is thus a sensitive test for density
{\it biasing}. It implies, subject to the errors,
that the $\sim\!12\hmpc$-smoothed
density fields of galaxies and mass are related via a biasing relation
that could be crudely approximated by a linear relation
with $b$ of order unity (but see a refinement of this in
\se{de_beta_biasing}).

Now that the data of peculiar velocities and the data of galaxy density
are found to be compatible with the model of \gi\ and linear biasing,
they can be combined to constrain
the degenerate parameter $\beta\equiv \Omega^{0.6}/b$.
The comparison between the two data sets
can be done in several different ways. In particular, it could be done
by comparing density fields derived locally
from the two data sets (\se{de_beta_dd}),
or by comparing velocities derived from the two data sets
(\se{de_beta_velmod}).
It can be done successively by first recovering fields from each data set
and then combining them to obtain $\beta$, or by a simultaneous recovery of
fields and beta determination from the two data sets
(\se{de_beta_simpot}).
It can be done by direct comparison of fields in r-space,
or by comparing coefficients mode by mode in a model expansion in z-space
(Davis, Nusser \& Willick 1996).

\subsection{Density-Density Comparison on Large Scales: POTIRAS}
\label{sec:de_beta_dd}

The main advantage of comparing densities is that they are {\it local}.
The densities are independent of long-range effects due to the unknown mass 
distribution outside the sampled volume, which could affect the velocities.
The densities are also independent of reference frame, and
can be reasonably corrected for non-linear effects.

The POTENT analysis extracts from the peculiar velocity data
a mildly-nonlinear mass density fluctuation field in a spatial grid, 
smoothed G12 (\se{de_rec_pot_veldel}).
The associated real-space density field of galaxies can be extracted 
with similar smoothing from a whole-sky redshift survey such as the 
IRAS 1.2 Jy survey (see Strauss \& Willick 1995; Sigad \etal 1997). 
 
A brief summary of the recovery of the IRAS density field is as follows.
The solution to the linearized \gi\ equation $\divv = -f\delta$
for an irrotational field is
\be
\vv(\vx)={f\over 4\pi} \int_{\rm all\ space}
d^3x'\, \delta(\vx') {\vx'-\vx\ \over \vert \vx'-\vx \vert^3} \ .
\label{eq:de_dipole1}
\ee
The velocity is proportional to the gravitational
acceleration, which ideally requires full knowledge of
the distribution of mass in space.
In practice, one is provided with a flux-limited, discrete redshift
survey, obeying some radial selection function $\phi(r)$.
The galaxy density is estimated by
$1\!+\!\delg(\vx)\!=\!\sum n^{-1}\phi(r_i)^{-1}\delta^3_{dirac}(\vx-\vx_i)$,
where $n\!\equiv\!V^{-1}\sum \phi(r_i)^{-1}$ is the mean galaxy density,
and the inverse weighting by $\phi$ restores the equal-volume weighting.
Eq.~\ref{eq:de_dipole1} is then replaced by
\be
\vv(\vx)={\beta\over 4\pi} \int_{\rm r<R_{max}}
d^3x'\, \delg(\vx')\, S(\vert\vx'-\vx\vert)\,
{\vx'-\vx\, \over \vert \vx'-\vx \vert^3} \ .
\label{eq:de_dipole2}
\ee
Under the assumption of linear biasing,
the cosmological dependence enters through $\beta$.
The integration is limited to $r\!<\!R_{max}$ where the signal dominates
over shot-noise.
$S(\vy)$ is a small-scale smoothing window ($\geq\! 500\kms$)
essential for reducing the effects of non-linear gravity,
shot-noise, distance uncertainty, and triple-value zones.
 
The distances are estimated from the redshifts in the LG frame by
\be
r_i = z_i -\hat{\vx}_i \cdot [\vv(\vx_i) - \vv(0)] \ .
\label{eq:de_ztor}
\ee
Equations~\ref{eq:de_dipole2}-\ref{eq:de_ztor}
can be solved iteratively: make a first guess for the
$\vx_i$, compute the $\vv_i$ by Eq.~\ref{eq:de_dipole2}, 
correct the $\vx_i$ by Eq.~\ref{eq:de_ztor}, and so on until convergence.  
The convergence can be improved by increasing $\beta$ gradually during the
iterations from zero to its desired value.

Even under $12\hmpc$ smoothing, $\delg$
is of order unity in places, necessitating a mildly-nonlinear treatment.
Local approximations from $\vv$ to $\delta$ were discussed in
\se{de_rec_pot_veldel}, but the non-local nature of the inverse problem 
makes it less straightforward.  
A possible solution is to find an inverse relation of
the sort $\divv\!=\!F(\Omega,\delg)$, including 
non-linear gravity and non-linear biasing.  
This is a Poisson-like equation in which
$-\beta\delg(\vx)$ is replaced by $F(\vx)$, and since the
smoothed velocity field is still irrotational for mildly-nonlinear 
perturbations, it can be integrated analogously to Eq.~\ref{eq:de_dipole1}.
With smoothing of
$10\hmpc$ and $\beta=1$, the approximation based on an empirical inverse to
$\delta_c$ (\ref{eq:de_delc})
has an \rms\ error $<50\kms$.

In recent applications, the galaxy density field is recovered 
from the noisy \iras\ data via a Power-Preserving Filter (PPF, by A. Yahil,
described in Sigad \etal 1997) --- a modification of the Wiener Filter.
The PPF returns a field that is not far from the Wiener, most probable field,
but it makes the result more realistic by forcing the variance to be 
constant in space despite the fact that the errors vary. 

\begin{figure}[th]
\vspace{6.5truecm}
{\includegraphics{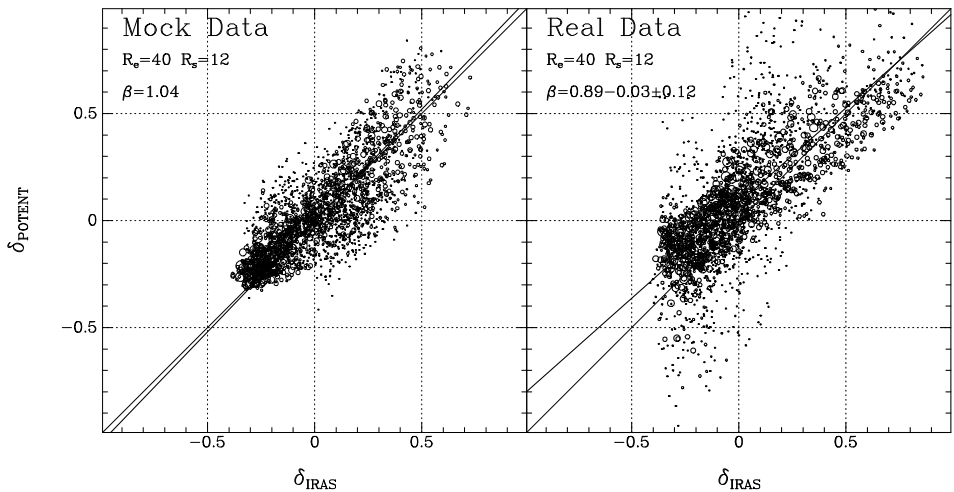}}
\caption{\capt 
$\beta$ from POTENT vs IRAS density comparison. The smoothing is G12
and the comparison volume is of effective radius $40\hmpc$.
Two-dimensional regression lines are marked.
Left: Regression between the averages of 20 mock catalogs of each type,
showing a bias as small as 4\%.
Right: Real data (Sigad \etal 1997).
}
\label{fig:de_potiras_scat}
\end{figure}

The simplest way of comparing the POTENT and IRAS density fields is via 
a two-dimensional linear regression using the values of the fields at 
grid points within a local comparison volume.  
The errors of both fields enter the regression.
The comparison volume is determined by equal-error contours. 

The latest comparison of POTENT Mark III and IRAS
1.2 Jy data at $12\hmpc$ smoothing within a volume of $(65 \hmpc)^3$ 
yields $\betai\!=\!0.86 \pm 0.12$ (Sigad \etal 1997).
The corresponding scatter diagram is shown in Figure \ref{fig:de_potiras_scat}.
The systematic error in this derivation, of only 4\%,
is deduced from the analogous
scatter diagram for the averages of 20 random mock catalogs of each type.
This is an update of the higher estimate $\betai\!=\!1.3\pm 0.3$ 
(Dekel \etal 1993) obtained based on an earlier version of POTENT with
the Mark II velocities and the IRAS 1.9 Jy redshifts.

 Similar comparisons of the mass density field with the density of
optical galaxies indicate a similar correlation and a
$\sim\!30\%$ lower estimate for $\betao$ (Hudson \etal 1995),
in agreement with the ratio of biasing factors, $\bo /\bi\!\approx\!1.3$,
obtained by direct comparison of optical and IRAS galaxy densities.

A comparison of similar nature of the POTENT Mark III data with the
density distribution of Abell/ACO $R\ge0$ clusters and the corresponding 
predicted velocities at G15 smoothing yields similar consistency out
to distances of $\sim 60\hmpc$, and an estimate of
$\betac=0.26 \pm 0.11$ (Plionis \etal 1997).
This is consistent with a linear biasing factor for the clusters that
is about 4 times larger than that of galaxies, in accordance with
the observed ratio of about $4^2$ for the corresponding correlation functions
(see Bahcall 1997).

\begin{figure}[th]
\vspace{6.9truecm}
{\includegraphics{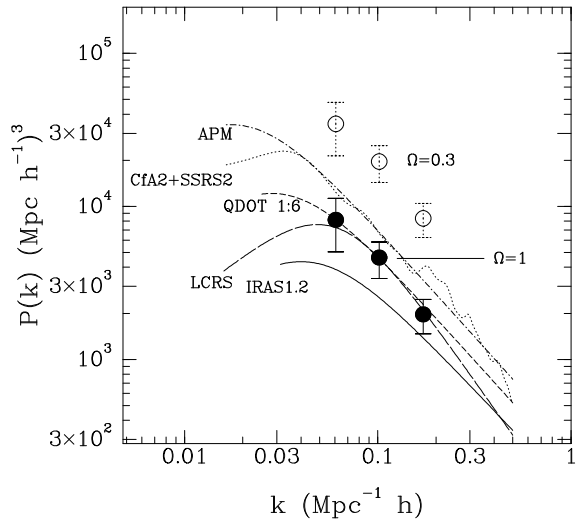}}
\caption{\capt 
$\beta$ from power spectra of galaxies versus mass.
The estimates from various galaxy density samples were all translated
from redshift to real space using Kaiser's approximation and
the best-fit value of $\beta$. The \pk\ from peculiar velocities
(Fig.~\ref{fig:de_spec}) is marked by
solid symbols for $\Omega=1$ (open symbols for $\Omega=0.3$). The values of
$\beta$ (for any $\Omega$) can be read directly from the vertical offset
of the solid symbols and the corresponding curves (Kolatt \& Dekel 1997).
}
\label{fig:de_beta_pk}
\end{figure}

A direct comparison of the mass power spectrum as
derived from peculiar velocities (\se{de_fluct_pot})
with the galaxy power spectra as derived from
different redshift and angular surveys is shown in Figure \ref{fig:de_beta_pk} 
(Kolatt \& Dekel 1997).  
It demonstrates a similarity in shape
and yields for the various galaxy types $\beta$ values in the range
$0.77-1.21$, with a typical error of $\pm 0.1$. For IRAS galaxies
typically $\beta \gsim 1$, and for optical galaxies $\beta \lsim 1$.
These estimates do not directly address the value of $\Omega$, but
it is clear from the figure that
if $\Omega$ is as small as $\sim\!0.3$, then all galaxy types must be
severely antibiased.

In principle, 
the degeneracy of $\Omega$ and $b$ is broken in the mildly-nonlinear
regime, where $\delta(\vv)$ is no longer $\propto\!f^{-1}$.  
Compatible mildly-nonlinear corrections in POTENT and in the IRAS analysis
allowed a preliminary attempt to separate these parameters using the
Mark II and IRAS 1.9 Jy data (Dekel \etal 1993).
Unfortunately, nonlinear biasing effects are hard to distinguish
from non-linear gravitational effects, so a specific nonlinear
biasing scheme is a prerequisite for such an analysis.

\subsection{Simultaneous Fit of Velocity and Density: SIMPOT}
\label{sec:de_beta_simpot}

The dynamical fields {\it and} $\beta$ can be recovered simultaneously
by a fit of a parametric model for the potential field
to the combined data of the observed radial peculiar velocities
and the distribution of galaxies in redshift space.
This procedure takes advantage of the complementary features of 
the data in the recovery of the fields, it enforces the same effective 
smoothing on the data without preliminary reconstruction procedures such
as POTENT, and it obtains a more reliable best fit by  
simultaneous rather than successive minimization. It has been implemented 
so far to the forward TF data, but it can be generalized in principle to
minimize inverse TF residuals.
 
In the SIMPOT procedure by Nusser \& Dekel (1997),
the model for the potential field is taken to be 
an expansion in spherical harmonics $Y_{lm}$ and Bessel functions $j_l$
where the coefficients $\Phi{lmn}$ are the free parameters,
\begin{equation}
\Phi(\vr )=
\sum_{l=1}^{l_{max}}\sum_{m=-l}^{l}\sum_{n=1}^{n_{max}
}\Phi_{lmn}\, j_l(k_nr) Y_{lm}(\hat{\vr}) .
\label{simpot_model}
\end{equation}
The model radial velocity is derived from this potential by 
$u=-\pa \Phi/\pa r$,
and the model density in redshift space is derived using linear theory via
$\delta=f^{-1} \nabla^2\Phi + s^{-2}(\pa/\pa s)(s^2\pa\Phi/\pa s)$.
The second term reflects redshift distortions, where $s$ is the radial 
variable in redshift space. The resulting models for $u(\vx)$ and $\delta(\vs)$
are expansions in certain functions $A_{lmn}$ and $B_{lmn}$ that are 
appropriate combinations of the original base functions. 
 
The combined $\chi^2$ to minimize as a function of the parameters 
$\Phi_{lmn}$ and $\beta$ is the sum of 
\begin{equation}
\chi^2_u=\sum_{i}\sigma_{ui}^{-2}
\left[u^{obs}_i-\sum_{lmn}A_{lmn}(\vr_i)\Phi_{lmn}\right]^2 
\label{eq:de_simpot_chiu}
\end{equation}
and
\begin{equation}
\chi^2_{\delta}=\int \dd^3s\ \sigma_{\delta}^{-2}(\vs)
\left[\delta^{obs}(\vs)-\sum_{lmn} B_{lnm}(\vs)\Phi_{lmn}\right]^2 .
\label{eq:de_simpot_chid}
\end{equation}
The observations are the peculiar velocities $u^{obs}_i$
and a continuous density field in redshift space $\delta^{obs}(\vs)$,
that is somewhat more tricky to obtain.
The $\beta$ dependence enters only in $\chi^2_\delta$, both via
the velocity-density relation and the redshift distortions.

A SIMPOT fit to the Mark III peculiar velocity data and
the IRAS 1.2 Jy redshift survey provides first hints for
scale-dependent biasing: 
$\betai\!\approx\! 0.6$ and $1.0$ ($\pm 0.1$)
for smoothings that are roughly equivalent
to Gaussian with radii $6$ and $12\hmpc$ respectively.

\subsection{Velocity-Velocity Comparison on Small Scales: VELMOD}
\label{sec:de_beta_velmod}

Earlier comparisons of the peculiar velocities from the Mark II catalog
and the velocities predicted from the IRAS redshift surveys (QDOT,
1.9 Jy, and 1.2 Jy) yielded estimates of $\betai$ in the range $0.4-1.0$
(Kaiser \etal 1991; Nusser \& Davis 1994).

The more sophisticated recent VELMOD method of comparison (Willick \etal 1997b)
compares the raw peculiar velocity data with a ``model" velocity field 
that is predicted from the IRAS 1.2 Jy redshift survey.
It is done without attempting to reconstruct a velocity field from the data.
The key feature of VELMOD is that it explicitly allows for a 
non-unique mapping between real space and redshift space.
Triple valuedness in the redshift field as
well as non-negligible small-scale velocity ``temperature" are treated
in a unified way. This is done by expressing the probability
that an object at distance $r$ has a redshift $z$ by
\begin{equation}
P(z|r) = {1\over \sqrt{2\pi}\sigma_u} \exp \left[ -{1\over 2}
{[z - r - u(r)]^2 \over 2\sigma_u^2}       \right]
\label{eq:de_velmod}
\end{equation}
where $u(r)$ is the radial component of the model velocity
field and $\sigma_u$ is the small-scale velocity noise
(which can in principle be a function of position).
The above probability is then multiplied by the TF probability
factor, $P(m,\eta,r)$, and integrated over the entire line-of-sight
to obtain the probability of the {\it observable} quantities
$(m,\eta,z)$.  One then maximizes that probability over the
entire data set.  

The method is computer-intensive because
numerical integrals are required for each galaxy, and for
each fit parameter (TF parameters, $\sigma_u$, velocity field
model parameters, etc.). This effort is worthwhile to the degree that
the velocity field is triple-valued or
the small-scale noise $\sigma_u$ is comparable to the TF error. 
In particular, VELMOD is more rigorous in an
analysis of the very local ($z\leq 3000\kms$) region.

VELMOD has been applied with a Gaussian smoothing of $3\hmpc$ 
to the IRAS 1.2 Jy redshift survey and a subset of 838 spiral galaxies 
from the Mark III catalog within $z\leq 3000\kms$ of the Local Group.
The method was tested successfully using mock catalogs drawn from the 
N-body simulation of Kolatt \etal (1996). 
When applied to the real data it
yielded consistency with the model of linear \gi\ and linear 
biasing once an artificial quadrupole was allowed, with 
$\betai\!=0.5\pm0.1$ at $3\hmpc$. The catch is that it is not at all clear
why linear \gi\ and the simplified deterministic biasing should be valid 
for the densities and velocities at such high resolution. 
The estimated value of $\beta$ should therefore be interpreted with caution.

It is interesting to note that Shaya \etal (1995) obtained a similarly low
value for $\betao$ from the same local neighborhood. They applied the 
least-action reconstruction method to a redshift survey
of several hundred spirals in comparison with TF data.  
Their method is likely to underestimate $\beta$ because it assumes that
the mass is all concentrated in the centers of galaxies and groups
and thus tends to overestimate the gravitational forces between them. 
This systematic effect is yet to be quantified.

\bigskip
Table~\ref{tab:de_beta}
summarizes the estimates of $\beta$ and $\Omega$ from cosmic flows.

\bigskip

\def\markiii{M3}
\def\markii{M2}
\def\IRAS{I}
\def\pote{$\scriptstyle{\rm}$}
 
\begin{table}
{
\baselineskip 12pt
\parskip 0pt
 
 
\def \srule {
        \vskip 0.4\baselineskip
        \hrule height.7pt
        \vskip 0.3\baselineskip }
\def \drule {
        \vskip 0.4\baselineskip
        \hrule height.7pt
        \vskip2pt
        \hrule height.7pt
        \vskip 0.3\baselineskip }
 
\caption{\capt $\Omega$ and $\beta$ from Cosmic Flows}
\label{tab:de_beta}
 
\vbox {
\drule
\halign to \hsize {
#\quad\hfil&#\quad\hfil&#\quad\hfil&#\hfil\cr
&&&\cr
 
Peculiar                       &Gaussian \ipdf\ \pote
 &Nusser \& Dekel 93           &$\Omega > 0.3 \ (>4\sigma)$ \cr
Velocities                     &Skewness($\divv$)\ \pote
 &Bernardeau \etal 94          &$\Omega > 0.3 \ (2\sigma)$       \cr
Alone                          &Void\ \pote
 &Dekel \& Rees 94             &$\Omega > 0.3 \ (2.4\sigma)$     \cr
                               &Power spectrum\ \pote
 &Kolatt \& Dekel 97        &$\sigma_8 \Omega^{0.6} = 0.7\pm0.15$\cr
                               &   +COBE
 &Zaroubi \etal 97a         & $\sigma_8 \Omega^{0.6} = 0.8\pm0.15$\cr
&&&\cr
\noalign{\srule}
&&&\cr
Galaxy                         &\markii-QDOT $v$
 &Kaiser \etal 91              &$\betai=0.9^{+0.2}_{-0.15}$      \cr
Density                       &\markiii-\IRAS1.2 $v$-dipole
 &Nusser \& Davis 94           &$\betai=0.6\pm0.2$               \cr
vs.                           &\markiii-\IRAS1.2 $v$-inverse
 &Davis \etal 96               &$\betai=0.6\pm0.2$(?)               \cr
Velocities                     &\markiii-\IRAS1.2 $v$ G3
 &Willick \etal 96             &$\betai = 0.5\pm0.1$        \cr
                            &\markiii-\IRAS1.2 $\delta$ G12
 &Sigad \etal 97               &$\betai=0.86\pm 0.15$                 \cr
                              &\markiii-\IRAS1.2 $\delta/v$ G6-12
 &Nusser \& Dekel 96           &$\betai=0.6-1.0$ scale  \cr
&&&\cr
                               &\markii-Optical $v$
 &Hudson 94                    &$\betao=0.5\pm0.1$               \cr
                               &TF-Optical
 &Shaya \etal 94               &$\betao=0.35\pm0.1$               \cr
                               &\markiii-Optical $\delta$ G12
 &Hudson \etal 95              &$\betao=0.75\pm0.2$               \cr
 
&&&\cr
                               &\markiii-clusters G15
 &Plionis \etal 97             &$\betac=0.26\pm0.11$             \cr
&&&\cr
\noalign{\srule}
&&&\cr
Redshift                       &$\xi$ \IRAS1.2
 &Peacock \& Dodds 94          &$\betai=1.0\pm0.2$               \cr
Distortions                    &$\xi$ \IRAS1.2
 &Fisher \etal 94a             &$\betai=0.45^{+0.3}_{-0.2}$ \cr
                               &$Y_{lm}$ \IRAS1.2
 &Fisher \etal 94b             &$\betai=1.0\pm0.3$   \cr
                               &$P_k$ \IRAS1.2, QDOT
 &Cole \etal 95                &$\betai = 0.5\pm 0.15$             \cr
                               &$\xi$ \IRAS1.2, QDOT
 &Hamilton 95                  &$\betai=0.7\pm 0.2$    \cr
                               &$Y_{lm}$ \IRAS1.2
 &Heavens \& Taylor 95         &$\betai=1.1\pm0.3$ \cr
                               &$P_k$ \IRAS1.2
 &Fisher \& Nusser 96          &$\betai=0.6\pm0.2$               \cr
&&&\cr
\noalign{\srule}
&&&\cr
\cmb\                         &vs galaxies angular
 &Yahil \etal 86             &$\betai=0.9\pm0.2$ \cr
Dipole                         &vs galaxies redshift
 &Strauss \etal 92            &$\betai=0.4-0.85$    \cr
                               &
 &Rowan-Rob. \etal 91     &$\betai=0.8^{+0.2}_{-0.15}$       \cr
&&&\cr
                              &vs galaxies angular
 &Lynden-Bell \etal 89        &$\betao=0.3-0.5$                 \cr
                               &vs galaxies redshift
 &Hudson 93                  &$\betao=0.7^{+0.4}_{-0.2}$         \cr
&&&\cr
                               & clusters
 &Scaramella \etal 91         &$\betac\sim0.13 $\cr
                              &
 &Plionis \etal 91            &$\betac\sim0.17-0.22$      \cr
&&&\cr
\noalign {\srule}
&&&\cr
}
}
\no$\beta\equiv\Omega^{0.6}/b$, \ \
               $\bc : \bo : \bi \approx 4.5 : 1.3 : 1.0$, \ \
               $\sigma_8\Omega^{0.6} = (0.69\pm0.05) \betai$, \\
\no \markiii = Mark III, \ \
               \IRAS1.2 = IRAS 1.2 Jy, \ \
               G12 = Gaussian smoothing $12\hmpc$
}
\end{table}
 

\subsection{Galaxy Biasing as a Stochastic Process}
\label{sec:de_beta_biasing}
 
\def\sb{\sigma_b}
\def\Sb{S_b}

In all the methods described in \se{de_beta}, the cosmological
parameter of interest $\Omega$ is contaminated by the uncertain relation
between galaxy and mass density, the so called ``galaxy biasing". 
Nontrivial galaxy biasing clearly exists.
The fact that galaxies of different types cluster differently (Dressler 1980)
implies that at least some do not trace the underlying mass.
This is hardly surprising because any reasonable physical theory 
would predict non-trivial biasing
(Kaiser 1984; Davis \etal 1985; Bardeen \etal 1986; Dekel \& Silk 1986; 
Dekel \& Rees 1987; Braun, Dekel \& Shapiro 1988; Weinberg 1995).
In particular, simulations of galaxy formation in a cosmological context
(\eg, Cen \& Ostriker 1992; 1993; Lemson \etal 1997)
indicate a biasing relation that is non-linear in density, is varying
with scale, and has a statistical scatter reflecting dependencies 
on factors other than density.
 
One should therefore not be surprised by the fact that
the various estimates of $\beta$ span a large range,
from less than one half to more than unity.
Some of this scatter is due to the different types of galaxies involved,
and some may be due to remaining effects of non-linear gravity or
other systematic errors, but a significant fraction of the scatter in $\beta$
is likely to reflect non-trivial properties of the biasing scheme.
This means that translating a measured $\beta$ into $\Omega$ is non-trivial;
it requires a detailed knowledge of the relevant biasing scheme.
 
In order to strengthen this point, we demonstrate below that
an obvious source of systematic variations in $\beta$ 
is the inevitable {\it statistical} scatter in the biasing process
(Dekel \& Lahav 1997).
This scatter in the relation between densities can be interpreted as
reflecting the dependence of galaxy formation efficiency, or galaxy density,
on physical properties of the protogalaxy environment other than density. 
These could be local properties such as the potential field,
the deformation tensor, tidal effects, and angular momentum,
or long-range effects carried by radiation or particles from 
neighboring sources. In the simple example below, we assume that this scatter
in the biasing is local and neglect possible spatial correlations.

Let $\delta (\vx)$ be the field of mass-density fluctuations smoothed 
with a given window,
and let $\delg(\vx)$ be the corresponding field for galaxies of a given type.
We treat them as random fields, both with probability densities
of zero mean by definition.
Denote $\langle\delta^2\rangle\equiv\sigma^2$ and 
$\langle\delta^3\rangle\equiv S$.
Consider the ``biasing" relation between galaxies and
mass to be a {\it random} process, specified by the
{\it conditional probability} function $B(\delg\vert\delta)$.
The common deterministic biasing relation, $\delg=b(\delta)\delta$,
is replaced by the conditional mean,
\begin{equation}
\langle\delg\vert\delta\rangle \equiv b(\delta)\delta .
\end{equation}
The statistical character of the relation is expressed by the conditional
moments of higher order about the mean, such as
\begin{equation}
\langle(\delg-b\delta)^2\vert\delta\rangle \equiv \sb^2, \quad  
{\rm and} \quad
\langle(\delg-b\delta)^3\vert\delta\rangle \equiv \Sb .
\end{equation}
This statistical nature of biasing leads to a different
``biasing parameter" for each specific application.
 
Take for example the ratio of
variances, $b_2^2 \equiv {\langle\delg^2\rangle /\langle\delta^2\rangle}$,
such as being obtained by a ratio of power spectra or two-point correlation
functions, or by comparing the mass function of clusters to the variance of
$\delg$ at $8\hmpc$ (White, Efstathiou \& Frenk 1993).
One can prove in general that
$\langle\delg^m\rangle = \langle\, 
\langle\delg^m\vert\delta\rangle_{\delg}\, \rangle_\delta $,
and therefore,
$ \langle\delg^2\rangle = 
\langle b^2(\delta)\,\delta^2\rangle + \langle\sb^2(\delta)\rangle $.
Thus, in the simple case where $b(\delta)$ is constant,
$b_2$ is an overestimate of $b$ by
\begin{equation}
b_2=b\, (1+\Delta_2)^{1/2},
\quad \Delta_2\equiv {\langle\sb^2\rangle /(\sigma^2 b^2)} .
\end{equation}
 
Another common way of estimating $\beta$ is via linear
regression of the noisy field $-\divv(\vx)$ [$\approx f(\Omega)\delta(\vx)$]
on $\delg(\vx)$ (\se{de_beta_dd}),
or via a regression of the corresponding velocities.
The slope of the forward regression of $\delg$ on $\delta$ is
$
b_f={\langle\delg\delta\rangle / \langle\delta^2\rangle},
$
and the slope of the inverse regression of $\delta$ on $\delg$ is
$
b_i^{-1}={\langle\delta\delg\rangle / \langle\delg^2\rangle}.
$
In the case where $b$ is constant, $b_f=b$, and $b_i$ is an overestimate,
\begin{equation}
b_i = b\,  (1+\Delta_2) .
\end{equation}
The promising method of estimating $\beta$ from large-scale redshift 
distortions measures yet a different
quantity.
It turns out that most methods for determining $\beta$ lead
similarly to an underestimate.
 
The level of the effect depends on the actual values of
$\Delta_2$ and similar parameters.
One way to estimate the natural biasing scatter at a given smoothing scale
is by investigating goodness of fit of the density fields of mass and light
and the model of deterministic biasing. By requiring that $\chi^2=1$ per
degree of freedom one can estimate the scatter needed in addition
to the known errors. For example, Hudson \etal (1995) estimated
for optical galaxies versus POTENT Mark III mass at $12\hmpc$ smoothing
$\sb\sim 0.15$, which corresponds to $\Delta_2 \sim 0.25$.
%
Alternatively, one can estimate $\sb$ from theoretical simulations.
For example, preliminary hydro simulations (Cen \& Ostriker 1993)
yield $\sb=0.25$ under $10\hmpc$ Gaussian smoothing, i.e. $\Delta_2\sim 0.4$.
If $b=\Omega=1$, then the $\beta$ values derived by the various methods
are expected to span the range $0.7\leq\beta\leq1$, and this is solely
due to the dispersion in the biasing relation.
A large skewness in $B(\delg\vert\delta)$ may stretch this range even further.
 
A relevant moral from the biasing uncertainty is that methods for
measuring $\Omega$ independent of density biasing (\se{de_omega}) 
are desirable. However, it has to be born in mind that the galaxies may
also be biased tracers of the {\it velocity} field of the matter.
Such a ``velocity biasing" would affect any attempt to extract dynamical 
information from large-scale velocities. The expected magnitude of 
the velocity biasing in the standard scenarios of structure formation
is a matter of debate, and even it's sign is unclear (\eg, Summers \etal
1995). 
Based on recent simulations it seems likely to be limited to a 
$\sim 10-20\%$ effect.

\section{COSMOLOGICAL PARAMETERS}
\label{sec:de_param}

\def\oml{\Omega_\Lambda}
\def\omb{\Omega_b}
\def\omn{\Omega_\nu}
\def\omt{\Omega_{tot}}
\def\ho{H_0}
\def\to{t_0}

The previous sections discussed measurements of the mass-density
parameter $\Omega$ (directly or via $\beta$)
from large-scale structure on scales $10-100\hmpc$. 
In this section we try to put these estimates in a wider perspective
(see Dekel, Burstein \& White 1997 for a review).

One very interesting large-scale constraint that has not been discussed here
is based on cluster abundance, that can be predicted for a Gaussian field
via the Press-Schechter formalism, and is quite insensitive
to the shape of the power spectrum. The current estimates are
$\sigma_8 \omm^{0.6} \simeq 0.5-0.6$ (White, Efstathiou \& Frenk 1993; 
Eke \etal 1996;
Mo \etal 1996). This is only slightly lower than the estimates of 
$\sigma_8 \omm^{0.6} \simeq 0.7-0.8$ from the power spectrum of the
peculiar velocity data (\se{de_fluct_pot}, \se{de_fluct_cobe}).
Note that this quantity is related to $\betai$ via 
$\sigma_8 \omm^{0.6} = \sigma_{8I} \betai$, where
$\sigma_{8I}$ is the rms fluctuation of IRAS galaxies in a top hat window
of $8\hmpc$. 
With the estimate from the IRAS 1.2 Jy survey of $\sigma_{8I}=0.69 \pm 0.05$ 
(Fisher \etal 1994a), the results from cluster abundance and from
the $\delta-\delta$ POTENT-IRAS comparison (\se{de_beta_dd}) are in
pleasant agreement.

Constraints from virialized systems such as galaxies and clusters
on smaller comoving scales of $1-10\hmpc$
(\eg, Primack 1997; Bahcall 1997; Peebles 1997)
typically yield low values of $\omm \sim 0.2-0.3$, but with
several loopholes.  
Most interesting among these is the constraint involving the baryonic
fraction in clusters from X-ray data
and the estimates of $\omb$ from the observed
deuterium abundance and the theory of big-bang nucleosynthesis.
With baryonic fraction in the middle of the observed range 
$f_b=(0.03-0.08)h^{-3/2}$ (White \etal 1993; White \& Fabian 1995),
and with the recent estimates of $\omb h^2 \simeq 0.025$ 
(Tytler \etal 1996; Burles \& Tytler 1996),
the current estimate is $\omm \simeq 0.5 h_{65}^{-1/2}$. 
This result favors a low value of $\omm$, but 
$\omm=1$ cannot be definitively excluded.

The global cosmological measures commonly involve combinations of the 
cosmological parameters, such as the mass $\Omega$ ($\equiv \Omega_m$)
and the contribution of the vacuum energy density $\oml$. 
Constraints in the $\omm-\oml$ plane
are displayed in Figure \ref{fig:de_omlam}, and briefly discussed below.

\subsubsection {Occam's Razor.}
\label{sec:de_param_occam}
The above working hypotheses, and the order by which more specific
models should be considered against observations, are guided by the
principle of Occam's Razor, \ie, by simplicity and robustness to initial
conditions. 
It is commonly assumed that the simplest model is the Einstein-deSitter
model, $\omm=1$ and $\oml=0$. One property that makes it robust is the
fact that $\omm$ remains constant
at all times with no need for fine tuning at the initial conditions.
The most natural extension according to the
generic model of inflation is a flat universe, $\omt=1$, where
$\omm$ can be smaller than unity but only at the expense of a
nonzero cosmological constant.

\subsubsection {Classical Tests of Geometry.}
\label{sec:de_param_geometry}
The parameter-dependent large-scale geometry of space-time
is reflected in the volume-redshift relation.
There are two classical versions of the tests that utilize this dependence:
magnitude versus redshift (or ``Hubble diagram")
and number density versus redshift.
The luminosity distance and the angular-diameter distance to a redshift $z$,
which enter these tests,
depend on $\omm$ and $\oml$.
At $z \sim 0.4$, these distances happen to be (to a good approximation)
a function of the combination $\omm-\oml$ (not $q_0$) (Perlmutter \etal 1996).

The main advantage of such tests is that they are
direct measures of global geometry.
Supernovae type Ia are the popular current candidate for a standard
candle, based on the assumption that
stellar processes are not likely to vary much in time.

The first 7 supernovae analyzed by Perlmutter \etal (1996)
at $z\sim 0.4$ yield $-0.3 < \omm-\oml < 2.5$ as the 90\% two-parameter
likelihood contour (Fig. \ref{fig:de_omlam}).
For a {\it flat} universe they find for each parameter
$\omm=0.94^{+0.34}_{-0.28}$, and
$\oml < 0.51$ (or equivalently $\omm > 0.49$) at 95\% confidence.

\subsubsection {Number Count of Quasar Lensing.}
\label{sec:de_param_lensing}
This is a promising new version of the classical number density test.
When $\oml$ is positive and comparable
to $\omm$, the universe should have
gone through a phase of slower expansion in the recent cosmological
past, which should be observed as an accumulation of objects at a specific
redshift of order unity.
In particular, it should be reflected in the observed
rate of lensing of high-redshift quasars by foreground galaxies
(Fukugita \etal 1990).
The contours of constant lensing probability in the $\omm-\oml$ plane
for $z_s \sim 2$ happen to almost coincide with the lines $\omm-\oml=const.$
The limits from lensing are thus similar in nature to the limits from SNe Ia.

This test shares all the advantages of direct geometrical measures.
The high redshifts involved bring about a unique sensitivity to $\oml$,
compared to the negligible effect that $\oml$ has on the structure observed
at $z\ll 1$.

From the failure to detect the accumulation of lenses, the current limit for a
{\it flat} model is $\oml < 0.66$ (or $\omm > 0.36$)
at 95\% confidence (Kochanek 1996) (Fig. \ref{fig:de_omlam}).

\subsubsection {Microwave Background Acoustic Peaks.}
\label{sec:de_param_cmb}
This test is expected
to provide the most stringent constraints on the cosmological parameters
within a decade.
The next generation of CMB satellites (MAP, to be launched by
NASA in 2001, and in particular Plank, scheduled by ESA for 2004)
are planned to
obtain a precision at $\sim 10$ arc-minute resolution
that will either rule out the current framework of \gi
for structure formation or
will measure the cosmological parameters to high precision.
Detailed evaluation of Plank shows that
nominal performance and expected foreground subtraction noise will
allow parameter estimation with the following accuracy 
(ignoring systematics):
$H_0 \pm 1\%$, $\omt \pm 0.005$, $\oml \pm 0.02$, $\omb \pm 2\%$.

\begin{figure}[th]
\vspace{8.15truecm}
{\includegraphics{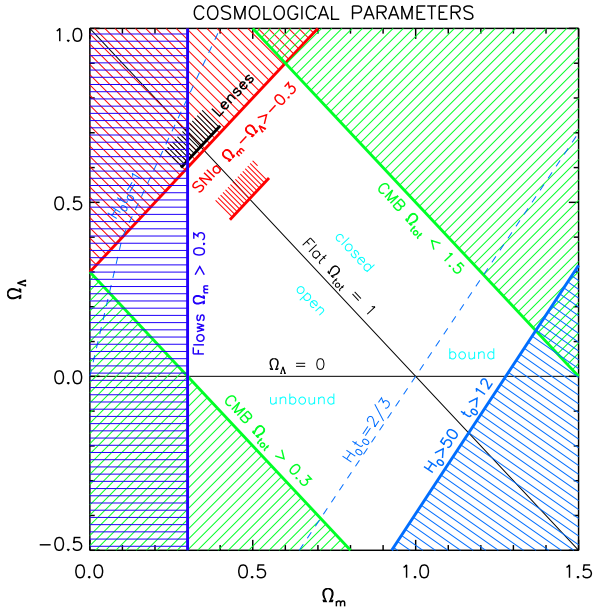}}
\caption{\capt 
Current limits ($\sim 2\sigma$)
on the cosmological parameters $\omm$ and $\oml$ from
global measures: luminosity distance of SNIa, lens count,
the location of the CMB peak, and the age versus Hubble constant.
The short marks are the one-parameter 95\% limits from SNIa and lenses
for a flat universe.
Also shown (vertical line) is the 95\% lower bound on $\omm$ from cosmic flows.
The most likely value of $\omm$ lies in the range 0.5 to 1.
The Einstein-deSitter model is permitted.
An open model with $\omm\simeq0.2$ and $\oml=0$,
or a flat model with $\omm\simeq 0.3$ and $\oml\simeq0.7$,
are ruled out (Dekel, Burstein \& White 1997).
}
\label{fig:de_omlam}
\end{figure}

Current ground-based and balloon-born experiments provide
preliminary constraints on the {\it location}
of the first acoustic peak on sub-degree scales in the angular power
spectrum of CMB temperature fluctuations, $l(l+1)C_l$.
In the vicinity of a flat model,
the first peak is predicted at approximately the multipole 
$ l_{peak} \simeq 220 (\omm+\oml)^{-1/2} $.
The results of COBE's DMR ($l\sim 10$) provide an upper bound of
$\omm+\oml < 1.5$ at the 95\% confidence level
for a scale-invariant initial spectrum
(and the constraint becomes tighter for any ``redder" spectrum, $n<1$)
(White \& Scott 1996).
Several balloon experiments ($l\sim 50-200$)
strengthen this upper bound (\eg, the Saskatoon experiment, Scott \etal 1996). 
The Saskatoon experiment and the CAT experiment ($l\sim 350-700$)
yield a preliminary lower bound of $\omm+\oml > 0.3$ (Hancock \etal 1996)
(Fig. \ref{fig:de_omlam}).

\subsubsection {The Age of the Universe.}
\label{sec:de_param_age}
Measured independent lower bounds on the Hubble constant and on the age
of the oldest globular clusters provide a lower bound on $\ho\to$
($=1.05 ht$, where $H_0 \equiv 100 h \kmsmpc$ and $t_0\equiv 10 t {\rm Gyr}$),
and thus an interesting constraint in the $\omm-\oml$ plane.
The exact expressions are computable in the various regions of
parameter space.
A useful crude approximation near $\ho\to \sim 2/3$ is
$
\omm -0.7\oml \simeq 5.8 (1- 1.3 ht) 
$.

Progress has been made in measuring $\ho$ via the HST key project
detecting Cepheids in nearby clusters for calibration of TF distances,
and via accurate distances to SNe Type Ia.
The new calibration of local Cepheids by the Hipparcos astrometric
satellite (Feast \& Catchpole 1997)
seem to have reduced the estimates of $\ho$ by $\sim 10\%$.
The indications from SNe velocities for a local void of radius 
$\sim 75\hmpc$ out to the Great Wall (Zehavi \etal 1997), 
takes another $\sim 5\%$
from $\ho$ as measured by TF distances from within the void,
and brings the various estimates into agreement at 
$h\simeq 0.6\pm 0.1$.

The Hipparcos calibration of the distances to local subdwarf stars
(Reid 1997) had an even more dramatic effect on the estimates of the ages
of the oldest globular clusters (\eg, Van den Berg \etal 1996).
The current estimates seem to be $t \simeq 1.2 \pm 0.2$ (\eg, M. Bolte, private
communication).
Thus, the most likely value of $\ho\to$ is not far from $2/3$, consistent
with the standard Einstein deSitter model.

\subsection{Conclusion}
\label{sec:de_conclusion}

We conclude with a summary of the main implications of the observed 
cosmic flows.

\subsubsection{Gravitational Instability.}
The strongest evidence for gravitational origin of structure comes from the
growth rate of fluctuations as indicated by the comparison of the
$\delta T/T\!\sim\!10^{-5}$ fluctuations at the last scattering surface
and the $\sim\!300\kms$ motions over $\sim\!100\hmpc$ scales in our
local neighborhood.

\subsubsection{Initial fluctuations and Dark Matter.} 
The COBE measurements of \cmb\ fluctuations at large angular scales
and the comparison to the observed flows indicate a power spectrum
near scale invariance, $n\!\sim\!1$. 

The bulk velocity in a sphere of radius $50\hmpc$ about the LG is
$V_{50}=375\pm85 \kms$.
The mass power spectrum deduced from the peculiar velocities has
an amplitude of
$P_{0.1} \Omega^{1.2} = (5\pm2)\times 10^3 (\!\hmpc)^3$
at $k=0.1 (\!\hmpc)^{-1}$.
This extrapolates to $\sigma_8 \Omega^{0.6} = 0.8 \pm 0.2$ on smaller scales.

For COBE-normalized CDM models, a likelihood analysis of the mass power
spectrum yields $\Omega n^2 h_{65} = 0.7 \pm 0.2$.
A comparison to preliminary detections of the first acoustic peak
in the \cmb\ angular power spectrum requires that $n\gsim 0.9$,
and that $\omb \sim 0.1$. 
Thus, within the family of CDM models, most successful in matching the 
current \lss\ data are either of the following variants:
(a) $\omm=1$ with a tilted spectrum $n\sim 0.9$,
(b) $\omm \sim 0.5$, with or without a cosmological constant, and $n=1$,
and
(c) $\omm=1$ with 20\% hot dark matter.
A high baryonic content (and relatively low $\omm$) may be required to
explain a peak in the galaxy density power spectrum at $\sim 125\hmpc$,
if confirmed
(Broadhurst \etal 1990; S. Landy based on LCRS, private communication; 
Cohen \etal 1996; Einasto 1997).

In view of the
nucleosynthesis constraints on the baryonic density, the high $\omm$
indicated by the motions requires non-baryonic dark matter.  
The observed mass-density
power spectrum on scales $10\!-\!100\hmpc$
does not yet allow a clear distinction between the competing models
involving baryonic, cold and hot dark matter
and possibly a cosmological constant.
I do not think that any of the
front-runner models can be significantly ruled out based on current tests,
contrary to occasional premature statements in the literature about the
``death" of a certain model.
I predict that were the dark matter constituent(s) to be securely detected in
the lab, the corresponding scenario of \lss\ would find a way to overcome
the $\sim\!2\sigma$ obstacles it may be facing now.

\subsubsection{Galaxy Biasing and $\beta$.}
Generally speaking, galaxies trace mass.
For each of the different smoothing scales, 
the data of velocities and redshift surveys are consistent with GI
and linear biasing (properly modified in the tails).
However, the best estimates of $\betai$ span the range $0.5-1.0$.
This can be explained by the fact that, when inspected in detail, 
the biasing scheme
involves scale dependence, non-linear features, and intrinsic scatter.
It is difficult to distinguish non-linear biasing from
non-linear gravitational effects. 

\subsubsection{The Value of $\omm$.}
Methods based on virialized objects tend to favor low values of 
$\Omega \sim 0.2$, but with plausible loopholes.
 
The current peculiar-velocity data provide in several different ways 
a significant ($>2\sigma$) lower bound of $\omm\!>\!0.3$.  
This bound is independent of $\Lambda$, $H_0$, and the biasing relation
between galaxies and mass.
The range of $\beta$ values obtained on different scales by different
methods may be partly due to underestimated errors and partly due to 
non-trivial biasing.

The global measures of geometry provide a lower bound of similar nature,
$\omm-\oml>0.3$.
The age constraints, which used to favor low values of $\omm$ until recently,
seem to agree with $\Omega\sim 1$ according to the new calibration of 
the distance scale by Hipparcos.

The data is thus consistent with $\omm\!=\!1$. 
Based on the whole range of constraints, and ignoring the Occam's razor
desire for simplicity, the most likely value may be argued to be
$\omm\!\sim\!0.5$. 
Values of $\omm=0.3$ and below are significantly ruled out.
The data are consistent with the general predictions of Inflation:
flat geometry and Gaussian, almost scale-invariant initial fluctuations.

\section*{Acknowledgments:}

This review is based on work with several close collaborators (see references),
supported by grants from the US-Israel Binational Science Foundation, 
the Israel Science Foundation, the NSF and NASA.

\bigskip\bigskip
 
 
\def\jo{}
\def\apj#1{{\jo ApJ,} {#1},~}
\def\apjl#1{{\jo ApJ,} {#1},~}
\def\apjs#1{{\jo ApJS,} {#1},~}
\def\mn#1{{\jo MNRAS,} {#1},~}
\def\aa#1{{\jo A\&A,} {#1},~}
\def\aj#1{{\jo AJ,} {#1},~}
\def\asp#1{{\jo PASP,} {#1},~}
\def\nat#1{{\jo Nature,} {#1},~}
\def\araa#1{{\jo ARA\&A,} {#1},~}
\def\baas#1{{\jo BAAS,} {#1},~}
\def\sc#1{{\jo Science,} {#1},~}
\def\pasj#1{{\jo PASJ,} {#1},~}
\def\rmp#1{{\jo RMP,} {#1},~}
\def\prl#1{{\jo PRL,} {#1},~}
\def\pl#1{{\jo Phys. Lett.,} {#1},~}
\def\prd#1{{\jo Phys. Rev. D.,} {#1},~}
\def\prb#1{{\jo Phys. Rev. B.,} {#1},~}
\def\phr#1{{\jo PhR,} {#1},~}
\def\al#1{{\jo Astrophys. Lett.,} {#1},~}
\def\ca#1{{\jo Comments Astrophys.,} {#1},~}
\def\asfiz#1{{\jo Astrofizika,} {#1},~}
\def\sa#1{{\jo Soviet Astron.,} {#1},~}
\def\rpp#1{{\jo Rep. Prog. Phys.,} {#1},~}
\def\anas#1{{\jo Ann. N. Y. Acad. Sci.,} {#1},~}
\def\pnas#1{{\jo Proc. Nat. Acad. Sci.,} {#1},~}
 
\def\jeru{in {\jo Formation of Structure in the Universe},
     eds. A. Dekel \& J.P. Ostriker (Cambridge Univ. Press)\ }


\def\bib#1{\bibitem{#1}}

\end{document}